\documentclass{iopart}
\usepackage{iopams}
\usepackage{setstack}
\def\sgn{\mathop{\rm sgn}\nolimits}
\def\res{\mathop{\rm res}}

\begin{document}
\title[On the extended resolvent of the Nonstationary Schr\"{o}dinger operator]{On the extended resolvent of the Nonstationary Schr\"{o}dinger operator for a Darboux transformed potential}
\author{M Boiti$^1$, F Pempinelli$^1$, and A K Pogrebkov$^2$}
\address{$^1$ Dipartimento di Fisica, Universit\`a
di Lecce and Sezione INFN, Lecce, Italy}
\address{$^2$ Steklov Mathematical Institute, Moscow, Russia}
\ead{boiti@le.infn.it, pempi@le.infn.it, pogreb@mi.ras.ru}

\begin{abstract}
In the framework of the resolvent approach it is introduced a so called
twisting operator that is able, at the same time, to superimpose \`a la
Darboux $N$ solitons to a generic smooth decaying potential of the
Nonstationary Schr\"odinger operator and to generate the corresponding Jost
solutions. This twisting operator is also used to construct an explicit
bilinear representation in terms of the Jost solutions of the related
extended resolvent. The main properties of the Jost and auxiliary Jost
solutions and of the resolvent are discussed.
\end{abstract}
\submitto{\JPA}
\maketitle

\section{Introduction}

The Kadomtsev--Petviashvili equation in its version called KPI~\cite{bib10}--\cite{bib12}
\begin{equation}
(u_{t}-6uu_{x_{1}}+u_{x_{1}x_{1}x_{1}})_{x_{1}}=3u_{x_{2}x_{2}},  \label{KPI}
\end{equation}
is a (2+1)-dimensional generalization of the celebrated Korteweg--de~Vries
(KdV) equation. As a consequence, the KPI equation admits solutions that
behave at space infinity like the solutions of the KdV equation. For
instance, if $u_{1}(t,x_{1})$ obeys KdV, then $u(t,x_{1},x_{2})=u_{1}(t,x_{1}+\mu x_{2}+3\mu ^{2}t)$ solves KPI for an
arbitrary constant $\mu \in \mathbb{R}$. Thus, it is natural to consider solutions
of~(\ref{KPI}) that are not decaying in all directions at space infinity but
have 1-dimensional rays with behaviour of the type of $u_{1}$. Even though
KPI has been known to be integrable for about three decades~\cite{bib11,bib12}, its general theory is far from being complete. Indeed, the
Cauchy problem for KPI with rapidly decaying initial data was resolved in~\cite{bib13}--\cite{bib16} by using the Inverse Scattering Transform (IST)
method on the base of the spectral analysis of the Nonstationary Schr\"{o}dinger operator
\begin{equation}
\mathcal{L}(x,i\partial _{x}^{{}})=i\partial _{x_{2}}^{{}}+\partial
_{x_{1}}^{2}-u(x),\qquad x=(x_{1}^{{}},x_{2}^{{}}),  \label{NS}
\end{equation}
that gives the associated linear problem for the KPI equation. However, it
is known that the standard approach to the spectral theory of the operator~(\ref{NS}), based on integral equations for the Jost solutions, fails for
potentials with one-dimensional asymptotic behaviour.

In~\cite{bib1}-\cite{bib9} the method of the ``extended resolvent'' (or, for
short, method of resolvent) was suggested as a way of pursuing a
generalization of the IST that enables studying the spectral theory of
operators with nontrivial asymptotic behaviour at space infinity. In~\cite{heat}--\cite{1-linemore} for the Nonstationary Schr\"{o}dinger and heat
operators the case where there is only one direction of nondecaying behaviour
was considered. The starting point in solving the problem was the embedding
of the pure one-dimensional case in the two-dimensional spectral theory,
building the two-dimensional extended resolvent for a potential $u(x)\equiv
u_{1}(x_{1})$. Then, a potential $u(x)=u_{1}(x_{1})+u_{2}(x)$, where $u_{2}(x)$ is an arbitrary decaying smooth function of both spatial variables, was considered and the corresponding resolvent was constructed by dressing the above resolvent for $u(x)=u_1(x_1)$. Finally, all mathematical entities generalizing the standard ones in IST, as Jost solutions and spectral data, were derived by a reduction procedure from this dressed resolvent.

Here, we consider the case of a potential not decaying along multiple non
parallel rays, which is substantially more complicated since we have
not the one-dimensional sample as a guide to follow and we must construct
directly the resolvent, without passing trough the embedding of
one-dimensional entities in two dimensions.

Therefore, we are obliged to consider directly true bidimensional
potentials. In \cite{backl-old}, by using recursively a binary Darboux
transformation \cite{matveev}, it was constructed explicitly not only a two
dimensional potential $\widetilde{u}$ which describes $N$ solitons \cite{KPsolitonsfirst} of the most general form \cite{generalKPsolitons}
``superimposed'' to a generic background but also its Jost solutions.
However, this recursive procedure seems not to be easily generalizable to
the construction of the corresponding extended resolvent. On the other side,
it is known (see~\cite{ams}) that, in the framework of the extended
resolvent approach, the whole hierarchy of time evolution equations related
to $\mathcal{L}$ in (\ref{NS}), which can be considered as infinitesimal
Darboux transformations, can be obtained by considering the similarity
transformation $\widetilde{L}=\zeta L\zeta ^{\dag }$ of the extended version
$L$ of $\mathcal{L}$, where $\zeta $ is a convenient unitary operator. Here,
we show that, by using a twisted transformation $\widetilde{L}\zeta =L\zeta
^{\dag }$ with $\zeta $ satisfying weaker conditions, one can bypass the
recursive procedure and build directly the final potential $\widetilde{u}$
and Jost solutions. Then, we use this operator $\zeta $, that we call
twisting operator, to build directly the extended resolvent of $\widetilde{L}
$ as a bilinear form in terms of the Jost solutions. The main properties of
the Jost and so called auxiliary Jost solutions and of the resolvent are
studied.

In a forthcoming paper, following the method developed in~\cite{bib9}--\cite{1-linemore}, we generalize the result obtained in this paper. Precisely, we
consider the potential obtained by adding to the potential describing $N$
solitons an arbitrary bidimensional smooth perturbation, we construct the
corresponding extended resolvent by dressing this one obtained in this paper
and we derive the corresponding Inverse Scattering problem.

\section{Background theory}

\label{background}

\subsection{Extended operators and resolvent}

In this section we briefly review the basic elements of the extended
resolvent approach. For further details, we refer the interested readers to~\cite{bib1}--\cite{heat}.

Let us consider the operators with kernel $L(x,x^{\prime })=\mathcal{L}(x,i\partial _{x}^{{}})\delta (x-x^{\prime })$ where $\mathcal{L}(x,i\partial _{x})$ denotes a differential operator whose coefficients are
smooth functions of $x$ and let us introduce what we call the \textbf{extension} of these differential operators, i.e., to any differential
operator $\mathcal{L}$ we associate the operator $L(\mathbf{q})$ with kernel
\begin{equation}
L(x,x^{\prime };\mathbf{q})\equiv \mathcal{L}(x,i\partial _{x}^{{}}+\mathbf{q})\delta (x-x^{\prime })=e_{{}}^{i\mathbf{q}(x-x^{\prime })}L(x,x^{\prime }),
\label{2}
\end{equation}
where $x=(x_{1},x_{2})$, $x^{\prime }=(x_{1}^{\prime },x_{2}^{\prime })\in \mathbb{R}^{2}$, and $\mathbf{q}=(\mathbf{q}_{1},\mathbf{q}_{2})\in \mathbb{C}^{2}$
and
\begin{equation*}
\mathbf{q}x=\mathbf{q}_{1}x_{1}+\mathbf{q}_{2}x_{2}.
\end{equation*}
The $\mathbf{q}$ variable will play in the following the role of a spectral
parameter and we use a bold face character to emphasize that it is complex.
By using the Fourier transform we can write
\begin{equation*}
L(x,x^{\prime };\mathbf{q})=\frac{1}{(2\pi )^{2}}\int d\alpha \,e^{-i\alpha
(x-x^{\prime })}\mathcal{L}(x,\alpha +\mathbf{q}),\quad \alpha =(\alpha
_{1},\alpha _{2}).
\end{equation*}
Then, it is natural to introduce more general operators $A(\mathbf{q})$ with
kernel\begin{equation}
A(x,x^{\prime };\mathbf{q})=\frac{1}{(2\pi )^{2}}\int d\alpha \,e^{-i\alpha
(x-x^{\prime })}\mathcal{P}(x,\alpha +\mathbf{q})  \label{Pseudo}
\end{equation}
obtained by considering not just a polynomial $\mathcal{L}(x,\mathbf{q})$ in
$\mathbf{q}$ but a tempered distribution $\mathcal{P}(x,\mathbf{q})$ of the
six real variables $x$, $\mathbf{q}_{\Re }$ and $\mathbf{q}_{\Im }$. Notice
that
\begin{equation}
A(x,x^{\prime };\mathbf{q})=e^{i\mathbf{q}_{\Re }(x-x^{\prime
})}A(x,x^{\prime };q),\quad \quad q\equiv \mathbf{q}_{\Im },\label{qcontinuation}
\end{equation}
and that $A(x,x^{\prime };q)$ belong to the space $\mathcal{S}^{\prime }$ of
tempered distributions of the six real variables $x$, $x^{\prime }$ and $q=(q_{1},q_{2})$. Since definition (\ref{Pseudo}), up to the introduction of
the spectral parameter $\mathbf{q}$ by shifting $\alpha ,$ coincides with
the definition of a pseudo-differential operator we shall call the operators
belonging to the space $\mathcal{S}^{\prime }$ extended pseudo-differential
operators, or by short operators, and $\mathcal{P}(x,\mathbf{q})$ their
symbol.

In the following it results often useful to use instead of the symbol $\mathcal{P}(x,\mathbf{q})$ its Fourier transform with respect to $x$, i.e.
\begin{equation*}
A(p;\mathbf{q})=\frac{1}{(2\pi )^{2}}\int dx e^{ipx}\mathcal{P}(x,\mathbf{q}),\quad \quad
p=(p_{1},p_{2}).
\end{equation*}
From (\ref{Pseudo}) and (\ref{qcontinuation}) it follows that $A(p;\mathbf{q})$ is related to $A(x,x^{\prime };q)$ by
\begin{equation}
A(p;\mathbf{q})=\frac{1}{(2\pi )^{2}}\int dx\int dx^{\prime }e^{i(p+\mathbf{q}_{\Re
})x-i\mathbf{q}_{\Re }x^{\prime }}A(x,x^{\prime };q).  \label{xp}
\end{equation}
Then, we consider $A(x,x^{\prime };q)$ and $A(p;\mathbf{q})$ as the representation
of the operator $A(q)$, respectively, in the $x$-space and in the $p$-space.
The inverse of (\ref{xp}) is given by
\begin{equation}
A(x,x^{\prime };\mathbf{q}_{\Im })=\frac{1}{(2\pi )^{2}}\int dp\int d\mathbf{q}_{\Re
}e^{-i(p+\mathbf{q}_{\Re })x+i\mathbf{q}_{\Re }x^{\prime }}A(p;\mathbf{q}).  \label{px}
\end{equation}
On the space of this operators we define the hermitian conjugation as
\begin{equation}
A^{\dag }(x,x^{\prime };q)=\overline{A(x^{\prime },x;-q)},\qquad A^{\dag }(p;\mathbf{q})=\overline{A(-p;\overline{\mathbf{q}}+p)},  \label{conj}
\end{equation}
in terms of kernels in $x$ or $p$-spaces. For generic operators $A(q)$ and $B(q)$ with kernels $A(x,x^{\prime };q)$ and $B(x,x^{\prime };q)$ we
introduce the standard composition law
\begin{equation}
(AB)(x,x^{\prime };q)=\int dx^{\prime \prime }\,A(x,x^{\prime \prime
};q)\,B(x^{\prime \prime },x^{\prime };q),  \label{3}
\end{equation}
if the integral exists in terms of distributions. In terms of kernels $A(p;\mathbf{q})$ and $B(p;\mathbf{q})$ this composition takes the form of a shifted convolution
\begin{equation}
(AB)(p;\mathbf{q})=\int dp^{\prime }A(p-p^{\prime };\mathbf{q}+p^{\prime })B(p^{\prime };\mathbf{q}).
\label{3p}
\end{equation}

An operator $A$ can have an inverse $A^{-1}$ in the sense of this
composition, i.e., such that $AA^{-1}=I$ or $A^{-1}A=I$, where $I$ is the
unity operator in $\mathcal{S}^{\prime }$, $I(x,x^{\prime };q)=\delta (x-x^{\prime
}) $, being $\delta (x)=\delta (x_{1})\delta (x_{2})$ the two-dimensional $\delta $-function (or $I(p;\mathbf{q})=\delta (p)$ in $p$-space).

Of course the two representations in the $x$ and in the $p$-space are
equivalent and, in principle, one could work always in one of them. However,
it results often convenient to pass from one representation to the other.
Thus, the $p$-space is more suitable to study analyticity properties, while
boundedness is more easily studied in the $x$-space.

The extension of the Nonstationary Schr\"{o}dinger operator (\ref{NS}) is
given by
\begin{equation}
L=L_{0}-U  \label{L}
\end{equation}
where in the $x$-space
\begin{equation}
\fl L_{0}(x,x^{\prime };q)=\left[ i(\partial _{x_{2}}+q_{2})+(\partial
_{x_{1}}+q_{1})^{2}\right] \delta (x-x^{\prime }), \quad U(x,x^{\prime
};q)=u(x)\delta (x-x^{\prime }).\label{Lu}
\end{equation}
The main object of our approach is the extended resolvent (or resolvent for
short) $M(q)$ of the operator $L(q)$, which is defined as the inverse of the
operator $L$, i.e.,
\begin{equation}
LM=ML=I.  \label{M}
\end{equation}
Here we omit to specify the additional conditions that guarantee uniqueness
of the extended resolvent as solution of~(\ref{M}), referring, say, to~\cite{1-line,1-linemore}.

For a real potential $u(x)$, as we always consider in the following, we have
\begin{equation}
L^{\dag}=L,\qquad M^{\dag}=M.  \label{dag}
\end{equation}

Now, let us consider in this section the case of a rapidly decaying potential $u(x)$ in (\ref{Lu}).

One of the main advantage of the resolvent approach is that the dressing
operators can be obtained directly from the resolvent by means of a
truncation and reduction procedure. Thus, here the dressing operators $\nu $
and $\omega $ are defined by
\begin{equation}
\fl \nu (p;\mathbf{q}_{1})=(ML_{0})(p;\mathbf{q})\Bigr|_{\mathbf{q}=\ell (\mathbf{q}_{1})},\qquad \omega (p;\mathbf{q}_{1})=(L_{0}M)(p;\mathbf{q})\Bigr|_{\mathbf{q}=\ell (\mathbf{q}_{1}+p_{1})-p},  \label{nuoM}
\end{equation}
where we introduced the special two-component vector
\begin{equation}
\ell (\mathbf{k})=(\mathbf{k},\mathbf{k}^{2}).  \label{ell}
\end{equation}
In fact these operators dress, in the sense proposed and developed by
Zakharov--Shabat~\cite{bib11}, the operator $L(q)$ and its resolvent $M(q)$.
More precisely, they admit the following bilinear representation in terms of
$\nu $ and $\omega $
\begin{equation}
L=\nu L_{0}\omega ,\qquad M=\nu M_{0}\omega ,  \label{LM}
\end{equation}
where $L_{0}$ and $M_{0}$ are the bare operators
\begin{equation}
L_{0}(p;\mathbf{q})=\delta (p)(\mathbf{q}_{2}-\mathbf{q}_{1}^{2}),\qquad M_{0}(p;\mathbf{q})=\frac{\delta
(p)}{\mathbf{q}_{2}-\mathbf{q}_{1}^{2}}.  \label{L0M0}
\end{equation}
Dressing operators are mutually adjoint
\begin{equation}
\nu ^{\dag }=\omega ,  \label{nudag}
\end{equation}
mutually inverse
\begin{eqnarray}
\omega \nu =I,  \label{scalar} \\
\nu \omega =I,  \label{compl}
\end{eqnarray}
and obey the equations
\begin{equation}
L\nu =\nu L_{0},\qquad \omega L=L_{0}\omega .  \label{Lnu}
\end{equation}

The kernels of these operators in $p$-space obey asymptotic
\begin{equation}
\lim_{\mathbf{q}_{1}\rightarrow \infty }\nu (p;\mathbf{q})=\delta (p),\qquad \lim_{\mathbf{q}_{1}\rightarrow \infty }\omega (p;\mathbf{q})=\delta (p),  \label{asymptnu}
\end{equation}
are independent of $\mathbf{q}_{2}$ and analytic functions of the variable
$\mathbf{q}_{1}$ in the upper and lower half planes.

Spectral data and the inverse problem can be formulated in this operatorial
approach. Here, we give not details but only the main formulae, that will be
useful in the following. Let $\nu ^{\pm }$ and $\omega ^{\pm }$ denote the
operators with kernels being the limiting values of $\nu (p;\mathbf{q})$ and $\omega
(p;\mathbf{q})$ at the real $\mathbf{q}_{1}$-axis from above and below
\begin{equation}
\nu ^{\pm }(p;\mathbf{q}_{1})=\nu (p;\mathbf{q}_{1\Re }\pm i0),\qquad \omega ^{\pm }(p;\mathbf{q}_{1})=\omega (p;\mathbf{q}_{1\Re }\pm i0).  \label{nuopm}
\end{equation}
Then we get the relations
\begin{equation}
\nu ^{\pm }=\nu ^{\mp }F^{\mp },\qquad \omega ^{\pm }=F^{\pm }\omega ^{\mp },
\label{nuo:ip}
\end{equation}
where we introduced the spectral data
\begin{equation}
F^{\pm }=\omega ^{\pm }\nu ^{\mp }.  \label{Fpm}
\end{equation}
By construction the kernels $F^{\pm }(p;\mathbf{q})$ depend on three real variables $p_{1}$, $p_{2}$, and $\mathbf{q}_{1\Re }$ and thanks to~(\ref{nudag})--(\ref{Lnu})
these operators obey
\begin{eqnarray}
L_{0}(q)F^{\pm }=F^{\pm }L_{0}(q),\quad \mbox{when}\quad q_{1}=0,
\label{Fprop1} \\
(F^{\pm })^{\dag }=F^{\pm },  \label{Fprop2} \\
F^{+}F^{-}=I.  \label{Fprop3}
\end{eqnarray}
By means of the last equality and~(\ref{asymptnu}) it can be shown that the
kernels $F^{\pm }(p;\mathbf{q})$ have the representation
\begin{equation}
F^{\pm }(p;\mathbf{q})=\delta (p)+\delta (p_{2}-p_{1}(p_{1}+2\mathbf{q}_{1\Re }))f^{\pm
}(p_{1}+\mathbf{q}_{1\Re },\mathbf{q}_{1\Re }),  \label{Ff}
\end{equation}
where the functions $f^{\pm }(p_{1},\mathbf{q}_{1\Re })$ depend only on two real
variables.

\subsection{Hat-kernels, Green's functions and Jost solutions}

Let us associate to any operator $A(q)$ with kernel $A(x,x^{\prime };q)$ its
``hat-kernel''
\begin{equation}
\widehat{A}(x,x^{\prime };q)=e_{{}}^{q(x-x^{\prime })}A(x,x^{\prime };q).
\label{7}
\end{equation}
If $A(q)$ is an extended differential operator $L(q)$ then this procedure is
the inverse of~(\ref{2}), i.e., $\widehat{L}(x,x^{\prime };q)=\mathcal{L}(x,x^{\prime })$, while for a generic operator the hat-kernel can continue
to depend on $q$. For any extended differential operator $L(q)$ and for any
(not necessary differential) operator $B(q)$ the following relations hold
\begin{eqnarray*}
(\widehat{LB})(x,x^{\prime };q)& =\mathcal{L}(x,\partial _{x}^{{}})\widehat{B}(x,x^{\prime };q), \\
(\widehat{BL})(x,x^{\prime };q)& =\mathcal{L}^{\mbox{\scriptsize d}}(x^{\prime
},\partial _{x^{\prime }}^{{}})\widehat{B}(x,x^{\prime };q),
\end{eqnarray*}
where $\mathcal{L}^{\mbox{\scriptsize d}}$ is the operator dual to $\mathcal{L}$. In
particular, by~(\ref{M}) we have that
\begin{equation*}
\mathcal{L}(x,\partial _{x})\widehat{M}(x,x^{\prime };q)=\mathcal{L}^{\mbox{\scriptsize d}}(x^{\prime
},\partial _{x^{\prime }})\widehat{M}(x,x^{\prime };q)=\delta (x-x^{\prime
}),
\end{equation*}
so that the hat-kernel of the resolvent defines a family of Green's function
depending on the bidimensional parameter $q$. In particular, the Jost
solutions are defined by means of the following Green's function depending
on a complex spectral parameter $\mathbf{k}$
\begin{equation}
\mathcal{G}(x,x^{\prime },\mathbf{k})=\widehat{M}(x,x^{\prime };\ell _{\Im }(\mathbf{k})),
\label{Gk}
\end{equation}
with the vector $\ell (\mathbf{k})$ defined as in~(\ref{ell}), while the
advanced/retarded solutions are defined by the Green's functions $\mathcal{G}_{\pm }(x,x^{\prime })$ obtained by the following limiting procedure
\begin{equation}
\mathcal{G}_{\pm }(x,x^{\prime })=\lim_{q_{2}\rightarrow \pm
0}\lim_{q_{1}\rightarrow 0}M(x,x^{\prime };q).  \label{Gpm}
\end{equation}

The Jost solution $\Phi (x,\mathbf{k})$ and its dual $\Psi (x,\mathbf{k})$ are defined
by
\begin{eqnarray}
\Phi (x,\mathbf{k})=\int dx^{\prime }e^{-i\ell (\mathbf{k})x^{\prime }}\mathcal{L}_{0}^{\mbox{\scriptsize d}}(x^{\prime },\partial _{x^{\prime }}^{{}})\mathcal{G}(x,x^{\prime },\mathbf{k})  \label{PhiG} \\
\Psi (x^{\prime },\mathbf{k})=\int dx\,e^{i\ell (\mathbf{k})x}\mathcal{L}_{0}(x,\partial
_{x}^{{}})\mathcal{G}(x,x^{\prime },\mathbf{k}),  \label{PsiG}
\end{eqnarray}
or by using definitions (\ref{nuoM}) in terms of the dressing operators by
\begin{equation}
\Phi (x,\mathbf{k})=e^{-i\ell (\mathbf{k})x}\chi (x,\mathbf{k}),\qquad \Psi (x,\mathbf{k})=e^{i\ell (\mathbf{k})x}\xi (x,\mathbf{k}),  \label{Jost}
\end{equation}
where
\begin{equation}
\fl \chi (x,\mathbf{q}_{1})=\int dp\,e^{-ipx}\nu (p;\mathbf{q}_{1}),\qquad \xi (x,\mathbf{q}_{1})=\int
dp\,e^{-ipx}\omega (p;\mathbf{q}_{1}-p_{1}).  \label{nuchi}
\end{equation}
The $\mathbf{q}_{2}$ independence of the kernels $\nu (p;\mathbf{q})$ and $\omega (p;\mathbf{q})$
implies, thanks to~(\ref{px}) and~(\ref{7}), that the corresponding
hat-kernels $\nu (x,x^{\prime };q)$ and $\omega (x,x^{\prime };q)$ are
independent of $q_{2}$ and are proportional to $\delta (x_{2}-x_{2}^{\prime
})$. Thus, instead of~(\ref{Jost}) and~(\ref{nuchi}) we get relations
\begin{equation}
\fl\Phi (x,\mathbf{q}_{1})=\int dy\,\widehat{\nu }(x,y;\mathbf{q}_{1\Im })e^{-i\ell (\mathbf{q}_{1})y},\qquad \Psi (x,\mathbf{q}_{1})=\int dy\,e^{i\ell (\mathbf{q}_{1})y}\widehat{\omega }(y,x;\mathbf{q}_{1\Im }),  \label{hat:chi}
\end{equation}

Writing~(\ref{M}) in the form $M=M_{0}+M_{0}UM$ one can derive from~(\ref{PhiG}) and~(\ref{PsiG}) the standard integral equation for the Jost
solution in the case of a smooth potential rapidly decaying at space
infinity
\begin{equation}
\Phi (x,\mathbf{k})=e^{-i\ell (\mathbf{k})x}+\int dx^{\prime }\mathcal{G}_{0}^{{}}(x,x^{\prime },\mathbf{k})u(x^{\prime })\Phi (x^{\prime },\mathbf{k}),  \label{4}
\end{equation}
where
\begin{equation*}
\mathcal{G}_{0}^{{}}(x,x^{\prime },\mathbf{k})=\frac{\mathop{\rm sgn}\nolimits
x_{2}^{{}}}{2\pi i}\int d\alpha \,\theta (\alpha \mathbf{k}_{\Im
}^{{}}x_{2}^{{}})\,e_{{}}^{i\ell (\alpha -\mathbf{k})(x-x^{\prime })},
\end{equation*}
is the Green's function, introduced in~\cite{Manakov}. In our approach this
Green's function follows directly from the second equality in~(\ref{L0M0}),
transformation~(\ref{px}), and definition~(\ref{Gk}). Solvability of this
integral equation under some small norm assumptions was proved in~\cite{bib15} and thanks to~(\ref{4}) it is easy to show that $\chi (x,\mathbf{k})$ has
the asymptotic behaviour
\begin{equation}
\lim_{ x \rightarrow \infty }\chi (x,\mathbf{k})=1.
\label{asymptx}
\end{equation}

Properties of the dressing operators lead to corresponding properties of the
Jost solutions. Thus~(\ref{nudag}) gives
\begin{equation}
\overline{\Phi (x,\mathbf{k})}=\Psi (x,\overline{\mathbf{k}}),\qquad \overline{\chi (x,\mathbf{k})}=\xi (x,\overline{\mathbf{k}}),\quad \mathbf{k}\in \mathbb{C}.  \label{Jostconj}
\end{equation}
Property~(\ref{asymptnu}) is equivalent to
\begin{equation}
\lim_{\mathbf{k}\rightarrow \infty }\chi (x,\mathbf{k})=1,\qquad \lim_{\mathbf{k}\rightarrow
\infty }\xi (x,\mathbf{k})=1,  \label{asymptk}
\end{equation}
relations~(\ref{scalar}) and~(\ref{compl}) can be considered as scalar
product and completeness of the Jost solutions:
\begin{eqnarray}
\frac{1}{2\pi }\int dx_{1}\Phi (x,\mathbf{k})\Psi (x,\mathbf{k}+\alpha )=\delta (\alpha
),\quad \mathbf{k}\in \mathbb{C},\quad \alpha \in \mathbb{R},  \label{scalarJost} \\
\frac{1}{2\pi }\int d\mathbf{k}_{\Re }\Phi (x,\mathbf{k})\Psi (y,\mathbf{k})\Bigr|_{x_{2}=y_{2}}=\delta (x_{1}-y_{1});  \label{complJost}
\end{eqnarray}
and equalities~(\ref{Lnu}) are equivalent to the Nonstationary Schr\"{o}dinger equation and its dual:
\begin{equation}
\mathcal{L}(x,\partial _{x})\Phi (x,\mathbf{k})=0,\qquad \mathcal{L}^{\mbox{\scriptsize d}}(x,\partial
_{x})\Psi (x,\mathbf{k})=0.  \label{NSE}
\end{equation}

The Jost solutions are analytic in the complex plane of the spectral
parameter $\mathbf{k}$, $\mathbf{k}_{\Im }\neq 0$. Using for the boundary values at the
real $\mathbf{k}$-axis notations of the type~(\ref{nuopm}) we get from~(\ref{nuo:ip}) and~(\ref{Ff}) the standard~\cite{Manakov} nonlocal Riemann--Hilbert
problem\begin{equation}
\Phi ^{\pm }(x,k)=\Phi ^{\mp }(x,k)+\int d\alpha \,\Phi ^{\mp }(x,\alpha
)f^{\mp }(\alpha ,k),\quad k\in \mathbb{R}.  \label{IPPhi}
\end{equation}

\section{Twisting transformation}

Here, we construct a two dimensional potential together with its Jost
solutions which describes $N$ solitons \cite{KPsolitonsfirst,generalKPsolitons} ``superimposed'' to a generic background
by using a twisted transformation $\widetilde{L}\zeta =\zeta L$ from the
extended differential operator $L$ in (\ref{L}) to a new operator $\widetilde{L}$ of the same form obtained by means of an operator $\zeta $
which is isometric but not necessarily unitary. The use of the twisting
operator $\zeta $ allows us, bypassing the usual procedure
consisting in applying recursively binary Darboux transformations, to get directly not only the Jost and auxiliary Jost solutions but also the extended resolvent, which results to be a bilinear expression in terms of these Jost and auxiliary Jost solutions.

\subsection{Properties of twisting operator $\protect\zeta $}

\label{Relation}

Let us consider twisting the operator $L$ in~(\ref{L}) to a new operator $\widetilde{L}$ of the same kind
\begin{equation}
\widetilde{L}=L_{0}-\widetilde{U},\qquad \widetilde{U}(x,x^{\prime };q)=\widetilde{u}(x)\delta (x-x^{\prime }),  \label{tildeL}
\end{equation}
by means of an operator $\zeta $ according to the formula
\begin{equation}
\widetilde{L}\zeta =\zeta L.  \label{Lzeta}
\end{equation}
We consider a potential $u(x)$ in $L$ which is real, smooth and rapidly
decaying at space infinity and we search for a $\zeta $ such that the new
potential $\widetilde{u}(x)$ is also real and smooth, while condition of
rapid decaying is not imposed. Notice that since $L$ and $\widetilde{L}$ are
both selfadjoint from (\ref{Lzeta}) it follows that
\begin{equation}
\zeta ^{\dag }\widetilde{L}=L\zeta ^{\dag }.  \label{Lzetadag}
\end{equation}
In addition we require that $\zeta $ obeys the conditions

\begin{enumerate}
\item \label{cond:isom} the operator $\zeta $ is isometric
\begin{equation}
\zeta ^{\dag }\zeta =I,  \label{zeta2}
\end{equation}
but not necessarily unitary;

\item \label{cond2} the kernel $\zeta (p;\mathbf{q})$ is independent of $\mathbf{q}_{2}$,
\begin{equation}
\zeta (p;\mathbf{q})=\zeta (p;\mathbf{q}_{1});  \label{zeta1}
\end{equation}

\item \label{cond:asympt} the kernel $\zeta (p;\mathbf{q})$ obeys the asymptotic
condition
\begin{equation}
\lim_{\mathbf{q}_{1}\rightarrow \infty }\zeta (p;\mathbf{q})=\delta (p).  \label{zeta3}
\end{equation}
\end{enumerate}

Then, a specific transformation $\zeta $ is chosen by fixing its analyticity
properties in $\mathbf{q}_{1}$.

Notice that because $\zeta $ is not unitary the operator
\begin{equation}
P=I-\zeta \zeta ^{\dag },  \label{P}
\end{equation}
is not zero and since it satisfies\begin{equation}
P^{\dag }=P  \label{P1}
\end{equation}
and, thanks to (\ref{zeta2}),\begin{equation}
P^{2}=P,  \label{P2}
\end{equation}
it is an orthogonal projector.

We fix the analyticity properties of $\zeta $ by requiring that it generates
not only the new potential $\widetilde{u}$ via (\ref{Lzeta}) but also the
new dressing operators $\widetilde{\nu }$ and $\widetilde{\omega }$. In
fact, taking into account~(\ref{Lnu}) we get by~(\ref{Lzeta}) and (\ref{Lzetadag})
\begin{equation*}
\widetilde{L}\zeta \nu =\zeta \nu L_{0},\quad \quad \omega \zeta ^{\dag }\widetilde{L}=L_{0}\omega \zeta ^{\dag },
\end{equation*}
and, therefore, the operators $\widetilde{\nu }$ and $\widetilde{\omega }$
defined by\begin{eqnarray}
 \widetilde{\nu }\tau =\zeta \nu ,  \label{tildenu1} \\
 \tau ^{\dag }\widetilde{\omega }=\omega \zeta ^{\dag },  \label{tildenu2}
\end{eqnarray}
for any operator $\tau $ commuting with $L_{0}$ satisfy the equations\begin{equation}
\widetilde{L}\widetilde{\nu }=\widetilde{\nu }L_{0},\qquad \widetilde{\omega
}\widetilde{L}=L_{0}\widetilde{\omega }.  \label{tLnu}
\end{equation}
analogous to the equations~(\ref{Lnu}) satisfied by dressing operators $\nu $
and $\omega $ of $L$. Since $\widetilde{\nu }$ and $\widetilde{\omega }$ are
are mutually adjoint\begin{equation}
\widetilde{\nu }^{\,\dag }=\widetilde{\omega },  \label{tnudag}
\end{equation}
we can limit ourself to consider $\widetilde{\nu }$ and say that $\widetilde{\nu }$ can be considered the new dressing operator of $\widetilde{L}$ if the
operator $\tau $ and consequently $\zeta $ is chosen in such a way that the
kernel of the $\widetilde{\nu }$ in $p$-space

\begin{enumerate}\setcounter{enumi}{3}
\item \label{cond3} independent of $\mathbf{q}_{2}$
\begin{equation}
\widetilde{\nu }(p;\mathbf{q})=\widetilde{\nu }(p;\mathbf{q}_{1}),
\end{equation}

\item \label{cond4} is an analytic function of the variable $\mathbf{q}_{1}$ in the
upper and lower half planes, continuous on the two sides of the real axis
and obeys the asymptotic behaviour
\begin{equation}
\lim_{\mathbf{q}_{1}\rightarrow \infty }\widetilde{\nu }(p;\mathbf{q})=\delta (p).
\label{tasymptnu}
\end{equation}
\end{enumerate}

Since $\nu $, $\omega $, and $\zeta $ by (\ref{cond2}), are independent of $\mathbf{q}_{2}$ we deduce from (\ref{cond3}) that also $\tau $ must be
independent of $\mathbf{q}_{2}$ and, then, since $\tau $ commutes with $L_{0} $, that its kernel has the form
\begin{equation}
\tau (p;\mathbf{q})=\delta (p)\tau (\mathbf{q}_{1}).  \label{tau1}
\end{equation}
Moreover, from (\ref{cond4}) we get that $\tau (\mathbf{q}_{1})$ has asymptotic
behaviour\begin{equation}
\lim_{\mathbf{q}_{1}\rightarrow \infty }\tau (\mathbf{q}_{1})=1.  \label{sigmasympt}
\end{equation}
Taking into account~(\ref{scalar}) and (\ref{zeta2}) we get from (\ref{tildenu1}) and (\ref{tildenu2}) the scalar product of the new dressing
operators
\begin{equation}
\widetilde{\omega }\widetilde{\nu }=T^{-1},  \label{tildesc}
\end{equation}
where we introduced the selfadjoint operator\begin{equation}
T=\tau \tau ^{\dag }  \label{t}
\end{equation}
with kernel\begin{equation}
T(p;\mathbf{q})=\delta (p)t(\mathbf{q}_{1})  \label{Tt}
\end{equation}
where thanks to the composition law~(\ref{3p}) and definition (\ref{conj})\begin{equation}
t(\mathbf{q}_{1})=\tau (\mathbf{q}_{1})\overline{\tau (\overline{\mathbf{q}}_{1})}.  \label{ttau}
\end{equation}
From (\ref{tildesc}), thanks to (\ref{cond4}), we deduce that $1/t(\mathbf{q}_{1})$ and therefore $1/\tau (\mathbf{q}_{1})$ must be chosen to be a function
analytic in the upper and lower half $\mathbf{q}_{1}$-planes and
continuous on the two sides of the real axis. Therefore, we assume that

\begin{enumerate}\setcounter{enumi}{5}
\item \label{tauprop}$\tau (\mathbf{q}_{1})$ is meromorphic in the upper and lower half planes without zeros and with a finite
number of poles and satisfies asymptotic (\ref{sigmasympt}).
\end{enumerate}

Consequently, $t(\mathbf{q}_{1})$, which plays the role of transmission coefficient,
is analytic in the upper and lower half $\mathbf{q}_{1}$-planes,
continuous on the two sides of the real axis with no zeros and with poles at
the poles of $\tau (\mathbf{q}_{1})$ and $\overline{\tau (\overline{\mathbf{q}}_{1})}$.

The completeness relation of the new dressing operators, which will play a
crucial role in the following, can be written in terms of $T$ and the
projection operator $P$. In fact from~(\ref{tildenu1}), (\ref{tildenu2}),
and~(\ref{P}) we get
\begin{equation}
\widetilde{\nu }T\widetilde{\omega }+P=I.  \label{tildecompl}
\end{equation}
It is also worth to mention that the operator $P$ annihilates the new dressing
operators. Indeed, from (\ref{tildesc}) and (\ref{tildecompl}) we have
\begin{equation}
P\widetilde{\nu }=\widetilde{\omega }P=0.  \label{Pnuo}
\end{equation}
Finally, by using the completeness relation (\ref{compl}) from (\ref{tildenu1}) we have
\begin{equation}
\zeta =\widetilde{\nu }\tau \omega  \label{zetanuom}
\end{equation}
and, once given an operator $\tau $ satisfying the above described
conditions, the analyticity properties of $\zeta $ are fixed and consequently
$\zeta $ itself. However, in determining the analyticity properties of $\zeta
$ we have not yet required that also the new dressing operators $\widetilde{\nu }$ and $\widetilde{\omega }$ satisfy non local Riemann-Hilbert problems
analogous to them satisfied by $\nu $ and $\omega $ in (\ref{nuo:ip}). This
will be done in the next section.

\subsection{Transformed continuous spectrum}

In order to describe the discontinuity at the real axis in the complex $\mathbf{q}_{1}$ plane of the twisting operator $\zeta $ and of the new
dressing operators and Jost solutions we use notations~(\ref{nuopm}) and
mention that for any operator $A$ by definition~(\ref{conj}) we have
\begin{equation}
(A^{\pm })^{\dag }=(A^{\dag })^{\mp }.  \label{conj:pm}
\end{equation}
From ~(\ref{tildenu1}) in the limits $q_{1}\rightarrow \pm 0$ we get\begin{equation}
\widetilde{\nu }^{\pm }\tau ^{\pm }=\zeta ^{\pm }\nu ^{\pm }.  \label{tnu:pm}
\end{equation}
Multiplying from the left by $(\zeta ^{\dag })^{\pm }$, thanks to~(\ref{zeta2}), we get
\begin{equation}
\nu ^{\pm }=(\zeta ^{\dag })^{\pm }\widetilde{\nu }^{\pm }\tau ^{\pm }
\label{nump}
\end{equation}
Inserting (\ref{nuo:ip}) into the r.h.s. of~(\ref{tnu:pm}) and then
inserting $\nu ^{\mp }$ from (\ref{nump}) in the obtained equation we get
\begin{equation*}
\widetilde{\nu }^{\pm }\tau ^{\pm }=\zeta ^{\pm }(\zeta ^{\dag })^{\mp }\widetilde{\nu }^{\mp }\tau ^{\mp }F^{\mp }.
\end{equation*}
If we impose that $\widetilde{\nu }$ satisfies a Riemann-Hilbert problem
analogous to that one in (\ref{nuo:ip}) satisfied by $\nu $ we deduce,
taking into account that according to (\ref{Pnuo}) $\widetilde{\nu }^{\mp }$
has a left annihilator $P^{\mp }$, that
\begin{equation*}
\zeta ^{\pm }(\zeta ^{\dag })^{\mp }=c_{1}I+c_{2}(I+P^{\mp
})=c_{1}I+c_{2}\zeta ^{\mp }(\zeta ^{\dag })^{\mp }
\end{equation*}
with $c_{1}$ and $c_{2}$ constant. Then, since $\zeta $ satisfies the
asymptotic property (\ref{zeta3}) thanks to (\ref{zeta2}) we derive\begin{equation}
\zeta ^{+}=\zeta ^{-},
\end{equation}
that is

\begin{enumerate}\setcounter{enumi}{6}
\item \label{zetacont}the twisting operator $\zeta $ is continuous on the
real axis of the $\mathbf{q}_{1}$ plane.
\end{enumerate}

Let us, now, introduce the operator
\begin{equation}
\widetilde{F}^{\pm }=\tau ^{\pm }F^{\pm }(\tau ^{\pm })^{\dag }.
\label{tFpm}
\end{equation}
Its kernel equals (see~(\ref{Ff}))
\begin{equation}
\fl\widetilde{F}^{\pm }(p;\mathbf{q})=\delta (p)|\tau ^{\pm }(\mathbf{q}_{1\Re })|^{2}+\delta
(p_{2}-p_{1}(p_{1}+2\mathbf{q}_{1\Re }))\widetilde{f}^{\pm }(p_{1}+\mathbf{q}_{1\Re },\mathbf{q}_{1\Re }),  \label{tFf}
\end{equation}
where
\begin{equation}
\widetilde{f}^{\pm }(p_{1},\mathbf{q}_{1\Re })=\tau ^{\pm }(p_{1})f^{\pm }(p_{1},\mathbf{q}_{1\Re })\overline{\tau ^{\pm }}(\mathbf{q}_{1\Re }).
\end{equation}
Thanks to properties of $\tau ^{\pm }$, the operators $\widetilde{F}^{\pm }$
have properties~(\ref{Fprop1}) and~(\ref{Fprop2}). Taking into account~(\ref{conj:pm}) and~(\ref{Tt}) we get finally
\begin{equation}
\widetilde{\nu }^{\pm }T^{\pm }=\widetilde{\nu }^{\mp }\widetilde{F}^{\mp
},\qquad T^{\pm }\widetilde{\omega }^{\pm }=\widetilde{F}^{\pm }\widetilde{\omega }^{\mp },  \label{tnuo:ip}
\end{equation}
where the second equation is obtained by conjugation. Comparison of (\ref{tnuo:ip}) with~(\ref{nuo:ip}) motivates the definition of $\widetilde{F}^{\pm }$ as the continuous spectrum of the new potential $\widetilde{u}$.
Again, thanks to ~(\ref{conj:pm}) and~(\ref{Tt}), we see that relations~(\ref{Fprop3}) and ~(\ref{Fpm}) are modified, respectively, as
\begin{equation}
(T^{\pm })^{-1}\widetilde{F}^{\pm }(T^{\mp })^{-1}\widetilde{F}^{\mp }=I,
\label{tFprop3}
\end{equation}
and as
\begin{equation}
\widetilde{\omega }^{\pm }\widetilde{\nu }^{\mp }=(T^{\pm })^{-1}\widetilde{F}^{\pm }(T^{\mp })^{-1}.  \label{tFpm'}
\end{equation}

In the case where the background potential $u(x)$ is identically equal to
zero the second term in~(\ref{Ff}) is absent and $F^{\pm }=I$. It is natural
to preserve this property for the new spectral data, so we impose
\begin{enumerate}\setcounter{enumi}{7}
\item \label{cond5} unitarity:
\begin{equation}
|\tau ^{\pm }(\mathbf{q}_{1\Re })|=1,  \label{tau:mod}
\end{equation}
\end{enumerate}
that in terms of operators means that
\begin{equation*}
(\tau ^{\pm })^{\dag }=\left( \tau ^{\pm }\right) ^{-1}.
\end{equation*}

In analogy with~(\ref{Jost}) and~(\ref{nuchi}) we define
\begin{eqnarray}
\fl \widetilde{\chi }(x,\mathbf{q}_{1})=\int dp\,e^{-ipx}\widetilde{\nu }(p;\mathbf{q}_1),&\qquad
\widetilde{\xi }(x,\mathbf{q}_{1})=\int dp\,e^{-ipx}\widetilde{\omega }(p;\mathbf{q}_{1}-p_{1}),  \label{tnuchi} \\
\fl \widetilde{\Phi }(x,\mathbf{k})=e^{-i\ell (\mathbf{k})x}\widetilde{\chi }(x,\mathbf{k}),&\qquad
\widetilde{\Psi }(x,\mathbf{k})=e^{i\ell (\mathbf{k})x}\widetilde{\xi }(x,\mathbf{k}),
\label{tildeJost}
\end{eqnarray}
where notation~(\ref{ell}) is used. These functions can be defined also by
relations generalizing~(\ref{hat:chi}). Indeed, taking into account that the
function $\tau (\mathbf{q}_{1})$ has no zeroes we have by~(\ref{tildenu1}) and~(\ref{tildenu2}) that $\widetilde{\nu }(p;\mathbf{q}_{1})=(\zeta \nu )(p;\mathbf{q})/\tau (\mathbf{q}_{1}) $ and $\widetilde{\omega }(p;\mathbf{q}_{1})=(\omega \zeta ^{\dag })(p;\mathbf{q})/\overline{\tau (p_{1}+\overline{\mathbf{q}}_{1})}$. Inserting this equalities in~(\ref{tnuchi}) we derive by~(\ref{xp}), (\ref{7}), and~(\ref{hat:chi}) that
\begin{eqnarray}
\widetilde{\Phi }(x,\mathbf{q}_{1})=\frac{1}{\tau (\mathbf{q}_{1})}\int dy\,\widehat{\zeta }(x,y;\mathbf{q}_{1\Im })\Phi (y,\mathbf{q}_{1}),  \label{Phiz} \\
\widetilde{\Psi }(x,\mathbf{q}_{1})=\frac{1}{\overline{\tau (\overline{\mathbf{q}}_{1})}}\int dy\,\Psi (y,\mathbf{q}_{1})\widehat{\zeta }^{\dag }(y,x;\mathbf{q}_{1\Im }).
\label{Psiz}
\end{eqnarray}

In what follows we prove that $\widetilde{\Phi }$ is the Jost solution of
operator $\widetilde{\mathcal{L}}$ and $\widetilde{\Psi }$ is the Jost solution of
the dual operator. Thanks to~(\ref{tnudag}) we have like in~(\ref{Jostconj})
\begin{equation}
\overline{\widetilde{\Phi }(x,\mathbf{k})}=\widetilde{\Psi }(x,\overline{\mathbf{k}}),\qquad \overline{\widetilde{\chi }(x,\mathbf{k})}=\widetilde{\xi }(x,\overline{\mathbf{k}}),\quad \mathbf{k}\in \mathbb{C}.  \label{tJostconj}
\end{equation}
Under condition~(\ref{cond4}) functions $\widetilde{\Phi }(x,\mathbf{k})$, $\widetilde{\Psi }(x,\mathbf{k})$, $\widetilde{\chi }(x,\mathbf{k})$, and $\widetilde{\xi }(x,\mathbf{k})$ are analytic with respect to the variable $\mathbf{k}\in \mathbb{C}$ with only
discontinuity at the real axis and functions $\widetilde{\chi }(x,\mathbf{k})$, $\widetilde{\xi }(x,\mathbf{k})$ obey asymptotic condition~(\ref{asymptk}). Eq.(\ref{tnuo:ip}) enables us to write down relations between boundary values $\widetilde{\Phi }^{\pm }(x,k)$ and $\widetilde{\Psi }^{\pm }(x,k)$ of these
functions on the real axis. Indeed, taking into account definition~(\ref{3p}) of composition we get by~(\ref{Tt}), (\ref{tFf}), and~(\ref{tau:mod})
\begin{equation}
\widetilde{\Phi }^{\pm }(x,k)t^{\pm }(k)=\widetilde{\Phi }^{\mp }(x,k)+\int
d\alpha \,\widetilde{\Phi }^{\mp }(x,\alpha )\widetilde{f}^{\mp }(\alpha
,k),\quad k\in \mathbb{R},  \label{tIPPhi}
\end{equation}
that modifies relation~(\ref{IPPhi}), valid in the case of rapidly decaying
potential $u(x)$.

\subsection{Twisting transformation of the resolvent}

\label{Resolvent}The operator $\zeta $, once obtained the transformed
operator $\widetilde{L}$, can also be used to get the corresponding
resolvent $\widetilde{M}=\widetilde{L}^{-1}$. From~(\ref{Lzeta}) and the
definition of the resolvent in~(\ref{M}) we have that $\widetilde{M}$ and $M$
are related by the same relation as $\widetilde{L}$ and $L$, i.e.,
\begin{equation}
\widetilde{M}\zeta =\zeta M.  \label{Mzeta}
\end{equation}
Multiplying~(\ref{Lzeta}) and~(\ref{Mzeta}) from the right by $\zeta ^{\dag
} $ and recalling the definition~(\ref{P}) of $P$ we get
\begin{equation}
\widetilde{L}=\zeta L\zeta ^{\dag }+L_{\Delta },\qquad \widetilde{M}=\zeta
M\zeta ^{\dag }+M_{\Delta }  \label{LMdelta}
\end{equation}
with
\begin{equation}
L_{\Delta }=\widetilde{L}P=P\widetilde{L},\qquad M_{\Delta }=\widetilde{M}P=P\widetilde{M},  \label{delta}
\end{equation}
where the second terms in both equations follow by hermitian conjugation.
Since $P$ is a projection operator we get directly
\begin{eqnarray}
 M_{\Delta }P=PM_{\Delta }=M_{\Delta },  \label{MPM} \\
 L_{\Delta }M_{\Delta }=M_{\Delta }L_{\Delta }=P.  \label{LMP}
\end{eqnarray}
Viceversa, for $M_{\Delta }$ satisfying~(\ref{MPM}) and~(\ref{LMP}) the
operator $\widetilde{M}$ in~(\ref{LMdelta}) is the resolvent of $\widetilde{L}$. In fact we have by~(\ref{zeta2}), (\ref{M}), and~(\ref{LMP})
\begin{equation}
\widetilde{L}\widetilde{M}=(\zeta L\zeta ^{\dag }+L_{\Delta })(\zeta M\zeta
^{\dag }+M_{\Delta })=I+\zeta L\zeta ^{\dag }M_{\Delta }+L_{\Delta }\zeta
M\zeta ^{\dag }.
\end{equation}
To prove that the last two terms are zero we, first, note that thanks to~(\ref{zeta2}) and~(\ref{P}) we have that
\begin{equation*}
P\zeta =\zeta ^{\dag }P=0,
\end{equation*}
and, then, that by~(\ref{delta}) $\zeta ^{\dag }M_{\Delta }=\zeta^\dag P \widetilde{M}=0$ and as well $L_{\Delta }\zeta=\widetilde{L}P\zeta=0$.
So
\begin{equation}
\widetilde{L}\widetilde{M}=\widetilde{M}\widetilde{L}=I,
\end{equation}
where the second equation is obtained by conjugation.

In conclusion, in order to obtain $M_{\Delta }$, and consequently $\widetilde{M}$ via (\ref{LMdelta}), one can, first, find the general
solution $X$ (belonging to our space of operators) of the system
\begin{equation}
PX=X,\qquad XP=X,  \label{X}
\end{equation}
and, then, $M_{\Delta }$ will be the special $X$ satisfying
\begin{equation}
L_{\Delta }X=XL_{\Delta }=P.  \label{LDMD}
\end{equation}

\subsection{Construction of operator $\protect\zeta $}

\label{construction}

In correspondence with condition (\ref{tauprop}) let $\tau(\mathbf{q}_1)$ have poles at $\mathbf{q}_1=\overline{\lambda}_1,\dots,\overline{\lambda}_N$ and let us suppose, for simplicity, that they are simple and that
\begin{eqnarray}
\lambda _{m}\neq \lambda _{n},\quad \lambda _{m}\neq \overline{\lambda _{n}},\quad\forall \,m\neq n\nonumber\\
\lambda _{n\Im }\neq 0,\quad \forall \,n.  \label{lambda}
\end{eqnarray}
From properties (\ref{tauprop}) and (\ref{cond5}) of $\tau $ by dispersion
relation we get that
\begin{equation}
\fl\tau (\mathbf{q}_{1})=\prod_{n=1}^{N}\left( \frac{\mathbf{q}_{1}-\lambda _{n}}{\mathbf{q}_{1}-\overline{\lambda }_{n}}\right) ^{\theta (-\mathbf{q}_{1\Im }\lambda _{n\Im
})}\equiv \prod_{n=1}^{N}\frac{\mathbf{q}_{1}-\lambda _{n\Re }+i|\lambda _{n\Im }|\sgn\mathbf{q}_{1\Im }}{\mathbf{q}_{1}-\overline{\lambda }_{n}},  \label{tau}
\end{equation}
For the residua of this function we have
\begin{equation}
\tau _{m}=\res_{\mathbf{q}_{1}=\overline{\lambda }_{m}}\tau (\mathbf{q}_{1})\equiv
-2i\lambda _{m\Im }\prod_{n=1,\,n\neq m}^{N}\left( \frac{\overline{\lambda }_{m}-\lambda _{n}}{\overline{\lambda }_{n}-\overline{\lambda }_{m}}\right)
^{\theta (\lambda _{m\Im }\lambda _{n\Im })}.  \label{tau-j}
\end{equation}
From (\ref{ttau}) we get for the transmission coefficient
\begin{equation}
\fl t(\mathbf{q}_{1})=\prod_{n=1}^{N}\left( \frac{\mathbf{q}_{1}-\overline{\lambda }_{n}}{\mathbf{q}_{1}-\lambda _{n}}\right) ^{\sgn(\mathbf{q}_{1\Im }\lambda _{n\Im })}\equiv
\prod_{n=1}^{N}\left( \frac{\mathbf{q}_{1}-\lambda _{n\Re }+i|\lambda _{n\Im }|}{\mathbf{q}_{1}-\lambda _{n\Re }-i|\lambda _{n\Im }|}\right) ^{\sgn(\mathbf{q}_{1\Im })}.
\label{t'}
\end{equation}
The residua at the poles at $\mathbf{q}_{1}=\lambda _{m}$ and $\mathbf{q}_{1}=\overline{\lambda }_{m}$ are given by
\begin{eqnarray}
t_{m}=\res_{\mathbf{q}_{1}=\lambda _{m}}t(\mathbf{q}_{1})\equiv 2i\lambda _{m\Im
}\prod_{n=1,\,n\neq m}^{N}\left( \frac{\lambda _{m}-\overline{\lambda }_{n}}{\lambda _{m}-\lambda _{n}}\right) ^{\sgn(\lambda _{m\Im }\lambda _{n\Im
})},  \label{t-j} \\
\res_{\mathbf{q}_{1}=\overline{\lambda }_{m}}t(\mathbf{q}_{1})=\overline{t}_{m}.
\label{t-j'}
\end{eqnarray}
An alternative expression is given by\begin{equation}
t_{m}=\tau (\lambda _{m})\overline{\tau }_{m}.  \label{ttau-r}
\end{equation}

Now, we have all we need for building the twisting operator $\zeta $. First,
we use~(\ref{compl}) to rewrite~(\ref{tildenu1}) as
\begin{equation}
\zeta =\widetilde{\nu }\tau \omega .
\end{equation}

Now, from the analyticity properties in $\mathbf{q}_{1}$ of $\widetilde{\nu }$
stated in (\ref{cond3}), (\ref{cond4}), of $\tau $ as given in (\ref{tau})
and of $\omega $, from the continuity of $\zeta $ on the real $\mathbf{q}_{1} $-axis stated in (\ref{zetacont}) and the asymptotic property of $\zeta
(p;\mathbf{q}_{1})$ we have that $\zeta (p;\mathbf{q})$ satisfies the following
integral representation\begin{equation}
\zeta (p;\mathbf{q})=\delta (p)+\sum_{n=1}^{N}\tau _{n}\int dp^{\prime }\,\frac{\widetilde{\nu }(p-p^{\prime };\overline{\lambda }_{n})\omega (p^{\prime };\overline{\lambda }_{n}-p_{1}^{\prime })}{\mathbf{q}_{1}+p_{1}^{\prime }-\overline{\lambda }_{n}}.  \label{z1}
\end{equation}
The kernel $\zeta (p;\mathbf{q})$ results to be piecewise analytic with
discontinuities along the lines $\mathbf{q}_{1\Im }=-\lambda _{n\Im }$ and
it results do depend only on the values of $\widetilde{\nu }(p;\mathbf{q}_{1})$ at $\mathbf{q}_{1}=\overline{\lambda }_{n}$. Therefore, in order to
get $\zeta $ it is sufficient to construct the new dressing operator $\widetilde{\nu }$ at the special values of poles of $\tau $.

The kernel of $\zeta $ in the $x$-space is given by~(\ref{xp}). Then, for
its hat-kernel (see~(\ref{7})) we get
\begin{eqnarray}
\fl\widehat{\zeta }(x,x^{\prime };q)=\delta (x-x^{\prime }) -i\sgn(x_{1}-x_{1}^{\prime })\delta (x_{2}-x_{2}^{\prime })  \nonumber \\
 \times \sum_{n=1}^{N}\tau _{n}\theta ((q_{1}+\lambda _{n\Im
})(x_{1}-x_{1}^{\prime }))\widetilde{\Phi }(x,\overline{\lambda }_{n})\Psi
(x^{\prime },\overline{\lambda }_{n}),  \label{expl1}
\end{eqnarray}
where we used notations~(\ref{Jost}), (\ref{nuchi}), (\ref{tnuchi}), and~(\ref{tildeJost}). By conjugation (see~(\ref{conj}), (\ref{Jostconj}), and~(\ref{tJostconj})) we derive
\begin{eqnarray}
\fl \widehat{\zeta }^{\dag }(x,x^{\prime };q)=\delta (x-x^{\prime }) -i\sgn(x_{1}-x_{1}^{\prime })\delta (x_{2}-x_{2}^{\prime }) \nonumber \\
 \times \sum_{n=1}^{N}\overline{\tau }_{n}\theta ((q_{1}-\lambda _{n\Im
})(x_{1}-x_{1}^{\prime }))\Phi (x,\lambda _{n})\widetilde{\Psi }(x^{\prime
},\lambda _{n}).  \label{expl2}
\end{eqnarray}
In order to complete the construction of $\zeta $ we have to impose the
isometry condition~(\ref{zeta2}). Using~(\ref{expl1}) and~(\ref{expl2}) we
get that this condition is equivalent to
\begin{eqnarray*}
\fl  \delta (x_{2}-x_{2}^{\prime })\sum_{n=1}^{N}\overline{\tau }_{n}\theta
((q_{1}-\lambda _{n\Im })(x_{1}-x_{1}^{\prime }))\Phi (x,\lambda _{n})\widetilde{\Psi }(x^{\prime },\lambda _{n}) \\
\fl\qquad{ } + \delta (x_{2}-x_{2}^{\prime })\sum_{n=1}^{N}\tau _{n}\theta
((q_{1}+\lambda _{n\Im })(x_{1}-x_{1}^{\prime }))\widetilde{\Phi }(x,\overline{\lambda }_{n})\Psi (x^{\prime },\overline{\lambda }_{n})\\
\fl{ } =i\sgn(x_{1}-x_{1}^{\prime })\delta (x_{2}-x_{2}^{\prime })\sum_{m,n=1}^{N}\overline{\tau }_{m}\tau _{n}\Phi (x,\lambda _{m})\Psi (x^{\prime },\overline{\lambda }_{n})\sgn(q_{1}-\lambda _{m\Im })\sgn(q_{1}+\lambda
_{n\Im })\\
\fl\qquad\times \int dy_{1}\theta ((q_{1}-\lambda _{m\Im })(x_{1}-y_{1}))\theta
((q_{1}+\lambda _{n\Im })(y_{1}-x_{1}^{\prime }))\widetilde{\Psi }(y,\lambda
_{m})\widetilde{\Phi }(y,\overline{\lambda }_{n})\Bigr|_{y_{2}=x_{2}}.
\end{eqnarray*}
Let us, now, introduce the matrix
\begin{equation}
\Theta (x)=\Biggl\Vert\overline{\tau }_{m}\tau
_{n}\int\limits_{x_{1}}^{(\lambda _{m}+\lambda _{n})_{\Im }\infty }dy_{1}
\widetilde{\Psi }(y,\lambda _{m})\widetilde{\Phi }(y,\overline{\lambda }_{n})\Bigr|_{y_{2}=x_{2}}\Biggr\Vert_{m,n=1}^{N},  \label{A}
\end{equation}
where the factor $(\lambda _{m}+\lambda _{n})_{\Im }$ in the limit of
integration defines the sign of the infinity. Thanks to~(\ref{tildeJost}) it
is easy to check that these integrals are well defined if $\widetilde{\xi }(x,\lambda _{m})$ and $\widetilde{\chi }(x,\overline{\lambda }_{n})$ are
bounded at space infinity, and that by~(\ref{tJostconj}) this matrix is
hermitian:
\begin{equation}
\overline{\Theta _{mn}(x)}=\Theta _{nm}(x).  \label{m4}
\end{equation}
Thus
\begin{equation}
\overline{\tau }_{m}\tau _{n}\widetilde{\Psi }(x,\lambda _{m})\widetilde{\Phi }(x,\overline{\lambda }_{n})=-\partial _{x_{1}}\Theta _{mn}(x),
\label{m5}
\end{equation}
and the above condition is simplified to
\begin{eqnarray*}
\fl \sum_{m=1}^{N}\theta ((q_{1}-\lambda _{m\Im })(x_{1}-x_{1}^{\prime }))\Phi
(x,\lambda _{m})\left[ \overline{\tau }_{m}\widetilde{\Psi }(x^{\prime
},\lambda _{m})-i\sum_{n=1}^{N}\Theta _{mn}(x^{\prime })\Psi (x^{\prime },\overline{\lambda }_{n})\right] \\
\fl{ } +\sum_{m=1}^{N}\theta ((q_{1}+\lambda _{m\Im })(x_{1}-x_{1}^{\prime }))
\left[ \tau _{m}\widetilde{\Phi }(x,\overline{\lambda }_{m})+i\sum_{n=1}^{N}\Phi (x,\lambda _{n})\Theta _{nm}(x)\right] \Psi (x^{\prime },\overline{\lambda }_{m})=0,
\end{eqnarray*}
where $x_{2}^{{}}=x_{2}^{\prime }$. The function in the l.h.s.\ is piecewise
constant with respect to $q_{1}$, so this condition is equivalent to the
equalities
\begin{eqnarray}
\widetilde{\Phi }(x,\overline{\lambda }_{m})=\frac{1}{i\tau _{m}}\sum_{n=1}^{N}\Phi (x,\lambda _{n})\Theta _{nm}(x),  \label{t-chi} \\
\widetilde{\Psi }(x,\lambda _{m})=\frac{i}{\overline{\tau }_{m}}\sum_{n=1}^{N}\Theta _{mn}(x)\Psi (x,\overline{\lambda }_{n}).  \label{t-xi}
\end{eqnarray}

By~(\ref{expl1}) in order to construct the hat kernel of the operator $\zeta
$ we have to determine the functions $\widetilde{\Phi }(\overline{\lambda }_{m})$ and $\widetilde{\Psi }(\lambda _{m})$. Thanks to~(\ref{t-chi}) and~(\ref{t-xi}) this means that we have to determine the matrix $\Theta _{mn}$.
For this sake we insert these equations in~(\ref{m5}), that gives
\begin{equation*}
\partial _{x_{1}}\Theta _{mn}(x)=-\sum_{m^{\prime },n^{\prime }=1}^{N}\Theta
_{mm^{\prime }}(x)\Psi (x,\overline{\lambda }_{m^{\prime }})\Phi (x,\lambda
_{n^{\prime }})\Theta _{n^{\prime }n}(x),
\end{equation*}
that under assumption of invertibility of the matrix $\Theta (x)$ can be
rewritten as
\begin{equation}
\partial _{x_{1}}\Theta _{mn}^{-1}(x)=\Psi (x,\overline{\lambda }_{m})\Phi
(x,\lambda _{n}).  \label{m7}
\end{equation}
Let us introduce in analogy with~(\ref{A}) the matrix
\begin{equation}
B_{mn}(x)=\int\limits_{-(\lambda _{m}+\lambda _{n})_{\Im }\infty
}^{x_{1}}dy_{1}\Psi (y,\overline{\lambda }_{m})\Phi (y,\lambda
_{n})\Bigr|_{y_{2}=x_{2}},  \label{b3}
\end{equation}
where again the limits of integration are uniquely determined by the
asymptotic behaviour of $\Phi (x,\overline{\lambda }_{m})$ and $\Psi
(x,\lambda _{n})$ given by~(\ref{Jost}) and~(\ref{asymptx}). Moreover,
\begin{equation*}
\lim_{x\rightarrow \infty }B_{mn}(x)e_{{}}^{-i(\overline{\ell (\lambda _{m})}-\ell (\lambda _{n}))x}=\frac{-i}{\overline{\lambda }_{m}-\lambda _{n}},
\end{equation*}
and by~(\ref{Jostconj}) this matrix is hermitian,
\begin{equation*}
\overline{B_{mn}(x)}=B_{nm}(x).
\end{equation*}

Now~(\ref{m7}) gives
\begin{equation}
\Theta _{mn}(x)=(B(x)+C)_{mn}^{-1},  \label{mab}
\end{equation}
where we introduced the matrix $C=\Vert c_{mn}\Vert _{m,n=1}^{N}$
with matrix elements independent of $x_{1}$.

Let $C_{\pm }$ denote the matrices constructed from $C$ as
\begin{equation}
C_{\pm }=\Vert c_{mn};\,m,n=1,\ldots ,N;\,\pm \lambda _{m\Im }>0,\,\pm
\lambda _{n\Im }>0\Vert ,  \label{3.11}
\end{equation}
and $C_{+}=I$ ($C_{-}=I$) if all $\lambda _{n\Im }$ are negative (positive).
In~\cite{backl-old} it was shown that if the matrix $C$ is hermitian and
obeys the positiveness condition
\begin{equation}
\pm C_{\pm }>0  \label{3.12}
\end{equation}
the determinant $\det (B(x)+C)$ has no zeros on the $x$-plane including
infinity (for the case of zero background potential this result was obtained
in~\cite{generalKPsolitons}). Thus, the matrix $\Theta (x)$ given by~(\ref{mab}) exists for any $x$ and the r.h.s.'s of~(\ref{t-chi}) and~(\ref{t-xi})
are given explicitly in terms of the Jost solutions of the background
potential $u(x)$. Such matrix $\Theta (x)$ is hermitian indeed, so that by~(\ref{Jostconj}) equalities~(\ref{t-chi}) and~(\ref{t-xi}) give a special
case of~(\ref{tJostconj}) for $\mathbf{k}=\overline{\lambda }_{m}$.

Finally, inserting~(\ref{t-chi}) in~(\ref{expl1}) and~(\ref{t-xi}) in~(\ref{expl2}) we get the hat-kernels of $\zeta $, $\zeta ^{\dag }$. Multiplying
them by $e^{-q(x-x^{\prime })}$ (see~(\ref{7})) we get kernels
\begin{eqnarray}
\fl\zeta (x,x^{\prime };q)=\delta (x-x^{\prime }) -\sgn(x_{1}-x_{1}^{\prime
})\delta (x_{2}-x_{2}^{\prime })e^{-q_{1}(x_{1}-x_{1}^{\prime })}
\nonumber \\
\times \sum_{m,n=1}^{N}\theta ((q_{1}+\lambda _{n\Im
})(x_{1}-x_{1}^{\prime }))\Phi (x,\lambda _{m})\Theta _{mn}(x)\Psi
(x^{\prime },\overline{\lambda }_{n}),  \label{z-expl} \\
\fl\zeta ^{\dag }(x,x^{\prime };q)=\delta (x-x^{\prime }) +\sgn(x_{1}-x_{1}^{\prime })\delta (x_{2}-x_{2}^{\prime
})e^{-q_{1}(x_{1}-x_{1}^{\prime })}  \nonumber \\
 \times \sum_{m,n=1}^{N}\theta ((q_{1}-\lambda _{m\Im
})(x_{1}-x_{1}^{\prime }))\Phi (x,\lambda _{m})\Theta _{mn}(x^{\prime })\Psi
(x^{\prime },\overline{\lambda }_{n}),  \label{zd-expl}
\end{eqnarray}
proving by~(\ref{Jostconj}) and~(\ref{m4}) that they are mutually conjugate
in the sense of~(\ref{conj}).

\subsection{Transformed dressing operators and Jost solutions}

\label{tJost}

By using the constructed twisting operator we can obtain the dressing operators by~(\ref{tildenu1}) and~(\ref{tildenu2}), while expressions for
functions introduced in~(\ref{tildeJost}) follow from~(\ref{Phiz}) and~(\ref{Psiz}). Indeed, inserting~(\ref{z-expl}) and~(\ref{zd-expl}) in these
relations, we get
\begin{eqnarray}
\widetilde{\Phi }(x,\mathbf{k})& =\frac{1}{\tau (\mathbf{k})}\left[ \Phi (x,\mathbf{k})-\sum_{m,n=1}^{N}\Phi (x,\lambda _{m})\Theta _{mn}(x)\beta _{n}(x,\mathbf{k})\right] ,  \label{t-phi} \\
\widetilde{\Psi }(x,\mathbf{k})& =\frac{1}{\overline{\tau (\overline{\mathbf{k}})}}\left[
\Psi (x,\mathbf{k})-\sum_{m,n=1}^{N}\overline{\beta _{m}(x,\overline{\mathbf{k}})}\Theta
_{mn}(x)\Psi (x,\overline{\lambda }_{n})\right] ,  \label{t-psi}
\end{eqnarray}
where we denoted
\begin{equation}
\beta _{n}(x,\mathbf{k})=\int\limits_{-(\mathbf{k}+\lambda _{n})_{\Im }\infty
}^{x_{1}}dy_{1}\Psi (y,\overline{\lambda }_{n})\Phi (y,\mathbf{k})\Bigr|_{y_{2}=x_{2}},  \label{beta}
\end{equation}
so that by~(\ref{b3})
\begin{equation*}
B_{mn}(x)=\beta _{m}(x,\lambda _{n})=\overline{\beta _{n}(x,\lambda _{m})}.
\end{equation*}

Thanks to the properties of the Jost solutions $\Phi $ and $\Psi $, the
functions $\beta _{n}(\mathbf{k})$ are analytic in the complex domain of $\mathbf{k}$ with
exception of the real axis and of a pole at $\mathbf{k}=\overline{\lambda }_{n}$,
where by~(\ref{scalarJost})
\begin{equation*}
\res_{\mathbf{k}=\overline{\lambda }_{n}}\beta _{n}(x,\mathbf{k})=i.
\end{equation*}
By~(\ref{tau-j}) this proves that $\widetilde{\Phi }(x,\overline{\lambda }_{m})$ ($\widetilde{\Psi }(x,\lambda _{m})$) given in~(\ref{t-chi})
(correspondingly,~(\ref{t-xi})) are values at $\mathbf{k}=\overline{\lambda }_{m}$
(at $\mathbf{k}=\lambda _{m}$) of~(\ref{t-phi}) (correspondingly,~(\ref{t-psi})).

Taking into account the well known property of the determinants of bordered
matrices
\begin{equation*}
\frac{1}{\det \Gamma _{n}^{{}}}\left|
\begin{array}{ll}
\Gamma _{n}^{{}} & \Gamma _{\ast ,n+1}^{{}} \\
\Gamma _{n+1,\ast }^{{}} & \gamma _{n+1,n+1}^{{}}\end{array}
\right| =\gamma _{n+1,n+1}^{{}}-\Gamma _{n+1,\ast }^{{}}\Gamma
_{n}^{-1}\Gamma _{\ast ,n+1}^{{}},
\end{equation*}
it is easy to see that~(\ref{t-phi}) is exactly the Jost solution
constructed in~\cite{backl-old} as result of $N$ successive ``binary'' B\"{a}cklund transformations. It is annulated by the operator
\begin{equation}
\widetilde{\mathcal{L}}(x,i\partial _{x}^{{}})=i\partial _{x_{2}}^{{}}+\partial
_{x_{1}}^{2}-\widetilde{u}(x),  \label{tNS}
\end{equation}
iff the matrix $C$ in~(\ref{mab}) is independent also on $x_{2}$.
Correspondingly, $\widetilde{\Psi }(x,\mathbf{k})$ is the Jost solution of the dual
operator. In~(\ref{tNS}) the new potential (cf.~(\ref{NS})) is given by
means of the well known~\cite{matveev} relation
\begin{equation}
\widetilde{u}(x)=u(x)+2\partial _{x_{1}}^{2}\log \det \Theta (x).
\label{tuu}
\end{equation}
We proved in~\cite{backl-old} that this potential is smooth, real and finite
for all $x$ under condition~(\ref{3.12}) and that it decays in all
directions on the $x$-plane with exception of a finite number of directions $x_{1}-2\lambda _{j\Re }x_{2}=$const, where it tends to a unidimensional
soliton. It was also proved there that $\widetilde{\chi }(x,\mathbf{k})$ (see~(\ref{tildeJost})) is a bounded function of its variables, analytic in $\mathbf{k}$ with
a discontinuity at the real axis, and obeying asymptotic condition~(\ref{asymptk}), while instead of~(\ref{asymptx}) we have
\begin{equation}
\lim_{x_{1}\rightarrow -\mathbf{k}_{\Im }\infty }\widetilde{\chi }(x,\mathbf{k})=1,\qquad
\lim_{x_{1}\rightarrow \mathbf{k}_{\Im }\infty }\widetilde{\chi }(x,\mathbf{k})=\frac{1}{t(\mathbf{k})}.  \label{tasymptx1}
\end{equation}
The Jost solutions given by~(\ref{t-phi}) and~(\ref{t-psi}) obviously obey~(\ref{tJostconj}). Then properties of $\widetilde{\xi }(x,\mathbf{k})$ are the same
up to the asymptotic
\begin{equation}
\lim_{x_{1}\rightarrow -\mathbf{k}_{\Im }\infty }\widetilde{\xi }(x,\mathbf{k})=\frac{1}{t(\mathbf{k})},\qquad \lim_{x_{1}\rightarrow \mathbf{k}_{\Im }\infty }\widetilde{\xi }(x,\mathbf{k})=1,  \label{tasymptx2}
\end{equation}
where we used that by ~(\ref{ttau}) $\overline{t(\overline{\mathbf{k}})}=t(\mathbf{k})$.

Summarizing, we have that kernels $\zeta (x,x^{\prime };q)$, $\zeta ^{\dag
}(x,x^{\prime };q)$ and $\widetilde{\nu }(p;\mathbf{q})$, $\widetilde{\omega }(p;\mathbf{q})$
(as given by~(\ref{tnuchi})) belong to the space $\mathcal{S}^{\prime }(\mathbb{R}^{6})$,
so they define operators in the sense of definitions given in Sec.~\ref{background}. Moreover, these operators obey all conditions~(\ref{cond2})--(\ref{cond4}) that were imposed. Asymptotic behaviour~(\ref{tasymptx1}) and~(\ref{tasymptx2}) shows that the values of the Jost solutions at the poles of
$t(\mathbf{k})$, like in the one-dimensional case, have special relevance. Thus in
addition to the values given in~(\ref{t-chi}) we have to consider $\Phi
(x,\lambda _{j})$. Thanks to~(\ref{beta}) and~(\ref{mab}) we get from~(\ref{t-phi})
\begin{equation*}
\widetilde{\Phi }(x,\lambda _{j})=\frac{1}{\tau (\lambda _{j})}\sum_{m,n=1}^{N}\Phi (x,\lambda _{m})\Theta _{mn}(x)c_{mj}.
\end{equation*}
then by~(\ref{t'}), (\ref{ttau-r}), and~(\ref{t-chi}), (\ref{t-xi})
\begin{eqnarray}
\widetilde{\Phi }(x,\lambda _{m})=\frac{i}{t_{m}}\sum_{n=1}^{N}\widetilde{\Phi }(x,\overline{\lambda }_{n})\tau _{n}c_{nm}\overline{\tau }_{m},
\label{tPhi} \\
\widetilde{\Psi }(x,\overline{\lambda }_{m})=\frac{1}{i\overline{t}_{m}}\sum_{n=1}^{N}\tau _{m}c_{mn}\overline{\tau }_{n}\widetilde{\Psi }(x,\lambda
_{n}).  \label{tPsi}
\end{eqnarray}
Using these equalities we get
\begin{equation}
\sum_{n=1}^{N}\Bigl\{t_{n}\widetilde{\Phi }(x,\lambda _{n})\widetilde{\Psi }(x^{\prime },\lambda _{n})+\overline{t}_{n}\widetilde{\Phi }(x,\overline{\lambda }_{n})\widetilde{\Psi }(x^{\prime },\overline{\lambda }_{n})\Bigr\}=0.  \label{zero}
\end{equation}
Relation~(\ref{tPhi}) together with~(\ref{tIPPhi}) and the first equality
in~(\ref{asymptk}) close the formulation of the inverse problem for the Jost
solution $\widetilde{\Phi }(x,\mathbf{k})$. Analogously,~(\ref{tPsi}), (\ref{tIPPhi}) and the second equality in~(\ref{asymptk}) give the inverse problem for
the dual Jost solution $\widetilde{\Psi }(x,\mathbf{k})$.

\subsection{Boundedness of $\protect\widetilde{\Phi }(x,\lambda_{n})$, $\protect \widetilde{\Phi }(x,\overline{\lambda }_{n})$ and their dual}

\label{bound}

Here we prove that the Jost solutions $\widetilde{\Phi }(x,\lambda
_{n})$, $\widetilde{\Phi }(x,\overline{\lambda }_{n})$ and $\widetilde{\Psi }(x,\lambda _{n})$, $\widetilde{\Psi }(x,\overline{\lambda }_{n})$ are
bounded when $x$ tends to infinity, while specific asymptotic behaviour
essentially depends on the direction of the limiting procedure. Let us
notice that $\widetilde{\Phi }(x,\lambda _{n})$ and $\widetilde{\Phi }(x,\overline{\lambda }_{n})$ can be written as
\begin{equation*}
\fl\widetilde{\Phi }(x,\lambda _{n})=e^{\ell _{\Im }(\lambda _{n})x-i\ell _{\Re
}(\lambda _{n})x}\widetilde{\chi }(x,\lambda _{n}),\quad \widetilde{\Phi }(x,\overline{\lambda }_{n})=e^{-\ell _{\Im }(\lambda _{n})x+i\ell _{\Re
}(\lambda _{n})x}\widetilde{\chi }(x,\overline{\lambda }_{n}),
\end{equation*}
where $\widetilde{\chi }(x,\lambda _{n})$ and $\widetilde{\chi }(x,\overline{\lambda }_{n})$ are known to be bounded. In the limit $x\rightarrow \infty $
along a direction of the $x$-plane some of the exponents $e^{\ell _{\Im
}(\lambda _{n})x}$ are increasing, some are decreasing or bounded. Taking
into account that new potential, Jost solutions, and spectral data by
construction are symmetric functions of $\lambda _{1},\ldots ,\lambda _{N}$,
we renumber them in a way that, say $e^{\ell _{\Im }(\lambda _{n})x}$ for $n=1,\ldots ,s-1$ are decreasing or constant and $e^{\ell _{\Im }(\lambda
_{n})x}$ for $n=s,\ldots ,N$ are increasing, where $s=1,\ldots ,N+1$ is a
number depending on the direction on the $x$-plane. Then $\widetilde{\Phi }(x,\lambda _{n})$ are decreasing for $n=s,\ldots ,N$ and $\widetilde{\Phi }(x,\overline{\lambda }_{n})$ are bounded when $n=1,\ldots ,s-1$. In order to
consider the complementary intervals we write~(\ref{tPhi}) for $m\geq s$ in
the form
\begin{equation*}
\sum_{n=s}^{N}\widetilde{\Phi }(x,\overline{\lambda }_{n})\tau _{n}c_{nm}=\frac{t_{m}}{i\overline{\tau }_{m}}\widetilde{\Phi }(x,\lambda
_{m})-\sum_{n=1}^{s-1}\widetilde{\Phi }(x,\overline{\lambda }_{n})\tau
_{n}c_{nm}.
\end{equation*}
Thanks to conditions~(\ref{3.12}) matrix $C$ cannot have zero eigenvalue.
Then the same is valid for the matrix $\Vert c_{mn}\Vert _{m,n=s}^{N}$ in
the l.h.s.\ of this equality. Taking into account that all terms in the
r.h.s.\ are bounded, we conclude that $\widetilde{\Phi }(x,\overline{\lambda
}_{n})$ are bounded also in the interval $n=s,\ldots ,N$ and then for all $n$. Now by~(\ref{tPhi}) the same is valid for $\widetilde{\Phi }(x,\lambda
_{n})$. Boundedness of $\widetilde{\Psi }(x,\lambda _{n})$, $\widetilde{\Psi
}(x,\overline{\lambda }_{n})$ follows by conjugation.

\section{Transformed resolvent}

\subsection{Operator $\protect P$ and completeness relation}

We already mentioned in Sec.~\ref{Relation} the essential role played by the
operator $P$. Here we derive an explicit expression for its kernel and
present its properties. Inserting~(\ref{expl1}) and~(\ref{expl2}) in~(\ref{P}) and using~(\ref{b3}) we get
\begin{eqnarray*}
\fl\widehat{P}(x,x^{\prime };q)= i\sgn(x_{1}^{{}}-x_{1}^{\prime })\delta
(x_{2}^{{}}-x_{2}^{\prime })\sum_{m=1}^{N}\tau _{m}\theta ((q_{1}+\lambda
_{m\Im })(x_{1}^{{}}-x_{1}^{\prime }))\\
 \times \widetilde{\Phi }(x,\overline{\lambda }_{m})\left[ \Psi (x^{\prime },\overline{\lambda }_{m})+i\sum_{n=1}^{N}\overline{\tau }_{n}\widetilde{\Psi }(x^{\prime },\lambda _{n})B_{mn}(x^{\prime })\right]\\
{ }+ i\sgn(x_{1}^{{}}-x_{1}^{\prime })\delta (x_{2}^{{}}-x_{2}^{\prime
})\sum_{n=1}^{N}\overline{\tau }_{n}\theta ((q_{1}-\lambda _{n\Im
})(x_{1}^{{}}-x_{1}^{\prime })) \\
\times \left[ \Phi (x,\lambda
_{n})-i\sum_{m=1}^{N}\tau _{m}\widetilde{\Phi }(x,\overline{\lambda }_{m})B_{mn}(x)\right] \widetilde{\Psi }(x^{\prime },\lambda _{n})
\end{eqnarray*}
By~(\ref{t-chi}) $\Phi (x,\lambda _{n})=i\sum_{m=1}^{N}\tau _{m}\widetilde{\Phi }(x,\overline{\lambda }_{m})($ $\Theta (x))_{mn}^{-1}$, so thanks to~(\ref{mab}) and~(\ref{tPhi})
\begin{equation*}
\Phi (x,\lambda _{n})-i\sum_{m=1}^{N}\tau _{m}\widetilde{\Phi }(x,\overline{\lambda }_{m})B_{mn}(x)=\frac{t_{n}}{\overline{\tau }_{n}}\widetilde{\Phi }(x,\lambda _{n}).
\end{equation*}
Thanks to this equality and its complex conjugate we get
\begin{eqnarray*}
\fl\widehat{P}(x,x^{\prime };q) =i\delta (x_{2}^{{}}-x_{2}^{\prime })\sgn(x_{1}^{{}}-x_{1}^{\prime })\\
 \times \sum_{n=1}^{N}\Bigl[t_{n}\theta ((q_{1}-\lambda _{n\Im
})(x_{1}^{{}}-x_{1}^{\prime }))\widetilde{\Phi }(x,\lambda _{n})\widetilde{\Psi }(x^{\prime },\lambda _{n}) \\
{ } +\overline{t}_{n}\theta ((q_{1}+\lambda _{n\Im })(x_{1}^{{}}-x_{1}^{\prime
}))\widetilde{\Phi }(x,\overline{\lambda }_{n})\widetilde{\Psi }(x^{\prime },\overline{\lambda }_{n})\Bigr].
\end{eqnarray*}

The form of this expression shows a discontinuity of the r.h.s.\ at $x_{1}^{{}}=x_{1}^{\prime }$, while thanks to~(\ref{zero}) the actual
discontinuity is absent. In order to exploit this fact directly we write
\begin{eqnarray*}
\fl\widehat{P}(x,x^{\prime };q)=i \delta (x_{2}^{{}}-x_{2}^{\prime })\theta
(q_{1})\\
\times \sum_{n=1}^{N}\Bigl\{t_{n}\Bigl[ \theta (\lambda _{n\Im })\theta
(|q_{1}|-|\lambda _{n\Im }|)\theta (x_{1}^{{}}-x_{1}^{\prime })+\theta
(-\lambda _{n\Im })\theta (x_{1}^{{}}-x_{1}^{\prime })\\
{ }-\theta (\lambda _{n\Im })\theta (|\lambda _{n\Im }|-|q_{1}|)\theta
(x_{1}^{\prime }-x_{1}^{{}})\Bigr]\widetilde{\Phi }(x,\lambda _{n})\widetilde{\Psi }(x^{\prime },\lambda _{n}) \\
{ }+\overline{t}_{n}\Bigl[ \theta (\lambda _{n\Im })\theta
(x_{1}^{{}}-x_{1}^{\prime })+\theta (-\lambda _{n\Im })\theta
(|q_{1}|-|\lambda _{n\Im }|)\theta (x_{1}^{{}}-x_{1}^{\prime })\\
{ }- \theta (-\lambda _{n\Im })\theta (|\lambda _{n\Im }|-|q_{1}|)\theta
(x_{1}^{\prime }-x_{1}^{{}})\Bigr]\widetilde{\Phi }(x,\overline{\lambda }_{n})\widetilde{\Psi }(x^{\prime },\overline{\lambda }_{n})\Bigr\}+\mbox{h.c.},
\end{eqnarray*}
where hermitian conjugation is understood in the sense of~(\ref{conj}), so
that
\begin{eqnarray*}
\fl\widehat{P}(x,x^{\prime };q) =i\delta (x_{2}^{{}}-x_{2}^{\prime })\sgn q_{1}\theta (q_{1}(x_{1}^{{}}-x_{1}^{\prime })) \\
\times \sum_{n=1}^{N}\Bigl\{t_{n}\widetilde{\Phi }(x,\lambda _{n})\widetilde{\Psi }(x^{\prime },\lambda _{n})+\overline{t}_{n}\widetilde{\Phi }(x,\overline{\lambda }_{n})\widetilde{\Psi }(x^{\prime },\overline{\lambda }_{n})\Bigr\}\\
{ }-i\delta (x_{2}^{{}}-x_{2}^{\prime })\sgn q_{1}\sum_{n=1}^{N}\theta
(|\lambda _{n\Im }|-|q_{1}|)\\
\times \Bigl\{t_{n}\theta (q_{1}\lambda _{n\Im })\widetilde{\Phi }(x,\lambda _{n})\widetilde{\Psi }(x^{\prime },\lambda _{n})+\overline{t}_{n}\theta (-q_{1}\lambda _{n\Im })\widetilde{\Phi }(x,\overline{\lambda }_{n})\widetilde{\Psi }(x^{\prime },\overline{\lambda }_{n})\Bigr\}.
\end{eqnarray*}
Again, thanks to~(\ref{zero}) the first term cancels out and we get
\begin{eqnarray}
\fl\widehat{P}(x,x^{\prime };q) =-i\delta (x_{2}^{{}}-x_{2}^{\prime })\sgn q_{1}\sum_{n=1}^{N}\theta (|\lambda _{n\Im }|-|q_{1}|)  \nonumber \\
\fl\qquad\qquad\times \Bigl\{t_{n}\theta (q_{1}\lambda _{n\Im })\widetilde{\Phi }(x,\lambda _{n})\widetilde{\Psi }(x^{\prime },\lambda _{n})+\overline{t}_{n}\theta (-q_{1}\lambda _{n\Im })\widetilde{\Phi }(x,\overline{\lambda }_{n})\widetilde{\Psi }(x^{\prime },\overline{\lambda }_{n})\Bigr\}.
\label{hatP+}
\end{eqnarray}
This suggests the introduction of $q_{1}$-dependent solutions of the Nonstationary Schr\"odinger equation
\begin{eqnarray}
\fl\widetilde{\Phi }_{n}(x,q_{1}) =\theta (|\lambda _{n\Im }|-|q_{1}|)\Bigl\{\theta (q_{1}\lambda _{n\Im })\widetilde{\Phi }(x,\lambda _{n})+\theta
(-q_{1}\lambda _{n\Im })\widetilde{\Phi }(x,\overline{\lambda }_{n})\Bigr\},
\label{tPhiq} \\
\fl\widetilde{\Psi }_{n}(x,q_{1}) =\theta (|\lambda _{n\Im }|-|q_{1}|)\Bigl\{\theta (q_{1}\lambda _{n\Im })\widetilde{\Psi }(x,\lambda _{n})+\theta
(-q_{1}\lambda _{n\Im })\widetilde{\Psi }(x,\overline{\lambda }_{n})\Bigr\},
\label{tPsiq}
\end{eqnarray}
which we call auxiliary Jost solutions.

Then we rewrite~(\ref{hatP+}) as
\begin{equation}
\widehat{P}(x,x^{\prime };q)=-i\sgn q_{1}\delta (x_{2}^{{}}-x_{2}^{\prime
})\sum_{n=1}^{N}\vartheta _{n}(q_{1})\widetilde{\Phi }_{n}(x,q_{1})\widetilde{\Psi }_{n}(x^{\prime },q_{1})  \label{hatP++}
\end{equation}
where\begin{equation}
\vartheta _{n}(q_{1})=\theta (q_{1}\lambda _{n\Im })t_{n}+\theta
(-q_{1}\lambda _{n\Im })\overline{t}_{n}.  \label{tq}
\end{equation}
Thus the operator $P$ results to be a sum of operators
that are different from zero in the intervals $[-|\lambda _{n\Im }|,|\lambda
_{n\Im }|]$ and continuous at $x_{1}=x_{1}^{\prime }$. In particular,
\begin{equation*}
P(q)=0\quad \mbox{for}\quad |q_{1}|>\max_{n}|\lambda _{n\Im }|.
\end{equation*}
Finally, inserting into (\ref{tildecompl}) equation (\ref{Tt}) and equations (\ref{tnuchi}), (\ref{tildeJost}) relating
the dressing operators to the Jost solutions and the expression obtained for
$P$ in (\ref{hatP++}) we get the completeness relation for the Jost solutions
\begin{eqnarray}
\fl\frac{1}{2\pi } \int d\alpha \,\widetilde{\Phi }(x;\alpha +iq_{1})t(\alpha
+iq_{1})\widetilde{\Psi }(x^{\prime };\alpha +iq_{1}) \nonumber \\
\fl\qquad\qquad{ } -i\sgn q_{1}\sum_{n=1}^{N}\vartheta _{n}(q_{1})\widetilde{\Phi }_{n}(x,q_{1})\widetilde{\Psi }_{n}(x^{\prime },q_{1})=\delta
(x_{1}-x_{1}^{\prime }),\quad \mbox{for }x_{2}=x_{2}^{\prime }.
\end{eqnarray}

\subsection{Properties of auxiliary Jost solutions}

Functions $\widetilde{\Phi }_{n}(x,q_{1})$ and $\widetilde{\Psi }_{n}(x,q_{1})$ by definitions~(\ref{tPhiq}),~(\ref{tPsiq}) obey the
nonstationary Schr\"{o}dinger equation and, correspondingly, its dual. Their
conjugation property
\begin{equation*}
\overline{\widetilde{\Phi }_{n}(x,q_{1})}=\widetilde{\Psi }_{n}(x,-q_{1}),
\end{equation*}
follows from~(\ref{tJostconj}). These functions are different from zero
on the interval $|q_{1}|<|\lambda _{n\Im }|$ only, that guaranties
boundedness of $e^{-q_{1}x_{1}}\widetilde{\Phi }_{n}(x,q_{1})$ and $e^{q_{1}x_{1}}\widetilde{\Psi }_{n}(x,q_{1})$ when $x_{1}$ goes to infinity.
As the result the kernel $P(x,x^{\prime };q)$ defined by~(\ref{7}) is
tempered distribution with respect to all its variables that proves that
operator $P(q)$ belongs to the space of operators under consideration. The
naturally appeared above piecewise dependence of functions $\widetilde{\Phi }_{n}(x,q_{1})$ and $\widetilde{\Psi }_{n}(x,q_{1})$ on $q_{1}$ remembers the
same, but nontrivial, dependence discovered in studying the perturbation of
the one-line potential in~\cite{1-line,1-linemore}. It is necessary to
mention that functions $\widetilde{\Phi }_{n}(x,q_{1})$ and $\widetilde{\Psi
}_{n}(x,q_{1})$ are discontinuous at $q_{1}=0$, while thanks to~(\ref{zero}) $P(q)$ is continuous at this point.

To study the discontinuity of these functions at $q_{1}=0$ we use
the standard notations (cf.~(\ref{nuopm})) for the right and left limit at $q_{1}=0$ and we write
\begin{eqnarray*}
\widetilde{\Phi }_{n}^{\pm }(x)& =\theta (\pm \lambda _{n\Im })\widetilde{\Phi }(x,\lambda _{n})+\theta (\mp \lambda _{n\Im })\widetilde{\Phi }(x,\overline{\lambda }_{n}), \\
\widetilde{\Psi }_{n}^{\pm }(x)& =\theta (\pm \lambda _{n\Im })\widetilde{\Psi }(x,\lambda _{n})+\theta (\mp \lambda _{n\Im })\widetilde{\Psi }(x,\overline{\lambda }_{n}),.
\end{eqnarray*}
To find relations between these limiting values we have to partially
invert relations in~(\ref{tPhi}) and~(\ref{tPsi}). Taking into account that
the whole our construction here is explicitly symmetric with respect to the $\lambda _{n}$'s, we can renumber them in a way that the first ones for $n=1,\dots ,s-1$ are in the upper half plane and the last ones for $n=s,\dots
,N$ are in the lower half plane. Correspondingly the matrix $C$ decomposes
as follows
\begin{equation}
C=\left(
\begin{array}{ll}
C_{+} & Z \\
Z^{\dag } & C_{-}
\end{array}
\right) ,  \label{c}
\end{equation}
where $C_{\pm }$ are the matrices, positive and negative definite,
introduced in~(\ref{3.12}), and where $Z$, $Z^{\dag }$, which are mutually
adjoint, are rectangular matrices just filling the $C$ matrix. By introducing the vectors
\begin{eqnarray*}
\Phi _{+}(\lambda ) &=&(\widetilde{\Phi }_{1}^{+},\dots ,\widetilde{\Phi }_{s-1}^{+})\quad \Phi _{-}(\lambda )=(\widetilde{\Phi }_{s}^{-},\dots ,\widetilde{\Phi }_{N}^{-}) \\
\Phi _{+}(\overline{\lambda }) &=&(\widetilde{\Phi }_{1}^{-},\dots ,\widetilde{\Phi }_{s-1}^{-})\quad \Phi _{-}(\overline{\lambda })=(\widetilde{\Phi }_{s}^{+},\dots ,\widetilde{\Phi }_{N}^{+})
\end{eqnarray*}
and the diagonal matrices
\begin{equation*}
\tau =\mbox{diag}\{\tau _{j}\},\qquad t=\mbox{diag}\{t_{j}\},
\end{equation*}
decomposed in the diagonal matrices $\tau _{\pm }$ and $t_{\pm }$ in analogy
to~(\ref{c}), equation (\ref{tPhi}) can be rewritten as
\begin{eqnarray*}
\Phi _{+}(\lambda )t_{+}=i\Phi _{+}(\overline{\lambda })\tau _{+}C_{+}\tau
_{+}^{\dag }+i\Phi _{-}(\overline{\lambda })\tau _{-}Z^{\dag }\tau
_{+}^{\dag }, \\
\Phi _{-}(\lambda )t_{-}=i\Phi _{+}(\overline{\lambda })\tau _{+}Z\tau
_{-}^{\dag }+i\Phi _{-}(\overline{\lambda })\tau _{-}C_{-}\tau _{-}^{\dag }.
\end{eqnarray*}
Solving the last equality for $\Phi _{-}(\overline{\lambda })$ and inserting
the result into the first one, we get
\begin{equation*}
\widetilde{\Phi }_{m}^{+}(x)=i\sum_{n=1}^{N}\widetilde{\Phi }_{n}^{-}(x)\vartheta _{n}^{+}c_{nm}^{\prime },
\end{equation*}
where the matrix $C^{\prime }=\Vert c^{\prime}_{mn}\Vert_{m,n=1}^{N}$ equals
\begin{equation*}
\fl C^{\prime }=\left(
\begin{array}{ll}
(t_{+}^{-1})^{\dag }\tau _{+} & 0 \\
0 & (\tau _{-}^{-1})^{\dag }
\end{array}
\right) \left(
\begin{array}{ll}
C_{+}-ZC_{-}^{-1}Z^{\dag } & \quad iZC_{-}^{-1} \\
-iC_{-}^{-1}Z^{\dag } & -C_{-}^{-1}
\end{array}
\right) \left(
\begin{array}{ll}
\tau _{+}^{\dag }t_{+}^{-1} & 0 \\
0 & \tau _{-}^{-1}
\end{array}
\right)
\end{equation*}
and $\vartheta _{n}^{+}$ is one of the limiting values of~(\ref{tq}):
\begin{equation*}
\vartheta _{n}^{\pm }=\theta (\pm \lambda _{n\Im })t_{n}+\theta (\mp \lambda
_{n\Im })\overline{t}_{n},
\end{equation*}
that in analogy with~(\ref{t-j}), (\ref{t-j'}) can also be defined as
\begin{equation*}
\vartheta _{n}^{\pm }=\res_{\mathbf{k}=\lambda _{n\Re }\mp i|\lambda _{n\Im }|}t(\mathbf{k}).
\end{equation*}
Thanks to conditions on matrix $C$ given in section~\ref{construction}, the
matrix $C^{\prime }$ is hermitian and positive. The first property is
obvious. In order to check the second one we write for an arbitrary vector $(v_{+},v_{-})$
\begin{eqnarray*}
\fl (v_{+}^{\dag },v_{-}^{\dag })\left(
\begin{array}{ll}
C_{+}-ZC_{-}^{-1}Z^{\dag } & \quad iZC_{-}^{-1} \\
-iC_{-}^{-1}Z^{\dag } & -C_{-}^{-1}
\end{array}
\right) \left(
\begin{array}{l}
v_{+} \\
v_{-}
\end{array}
\right) \\
{ } =v_{+}^{\dag }C_{+}v_{+}-[v_{-}-iZ^{\dag }v_{+}]^{\dag
}C_{-}^{-1}[v_{-}+iZ^{\dag }v_{+}],
\end{eqnarray*}
that proves that conditions~(\ref{3.12}) are equivalent to condition $C^{\prime }>0$.

Now relation for the functions $\widetilde{\Psi }_{n}^{\pm }(x)$ follows by
complex conjugation, taking into account that
\begin{equation*}
\overline{\widetilde{\Phi }_{n}^{\pm }(x)}=\widetilde{\Psi }_{n}^{\mp
}(x),\qquad \overline{\vartheta _{n}^{\pm }}=\vartheta _{n}^{\mp }.
\end{equation*}
Let us mention that the only dependence on the signs of $\lambda _{n\Im }$
enters in the functions $\tau (\mathbf{k})$ in~(\ref{tau}), and then in the
spectral data~(\ref{tFf}). All other objects of our constructions depend on $|\lambda _{n\Im }|$ only. This means that in the case of the zero background
potential $u(x)$, when both $f^{\pm }$ and $\widetilde{f}^{\pm }$ are
identically equal to zero, without loss of generality we can chose in
addition to~(\ref{lambda}) $\lambda _{n\Im }>0$ ($n=1,2,\ldots ,N$).

On the other side in the case of the nonzero background potential solutions
corresponding to $\lambda _{n}$ with opposite sign are different, as was
shown in~\cite{backl-old}.

Finally, let us mention that relation $P^{2}=P$ gives the scalar product
\begin{equation}
\int dx_{1}\,\widetilde{\Psi }_{m}(x,q_{1})\widetilde{\Phi }_{n}(x,q_{1})=\frac{i\delta _{mn}\theta (|\lambda _{m\Im }|-|q_{1}|)}{(\sgn q_{1})\vartheta _{m}(q_{1})},  \label{scalar+}
\end{equation}
that takes a more complicated form if we turn back to the original solutions.

\subsection{Operators $\protect L_{\Delta}$ and $\protect M_{\Delta}$}

In order to find $L_{\Delta }$ we can use the first equality in~(\ref{delta}) applying $\widetilde{L}$ to $P$. Then in terms of the hat-kernels we use
that $\widetilde{\Phi }_{m}^{\pm }(x)$ is annulated by $\widetilde{\mathcal{L}}$, so
\begin{equation}
\widehat{L}_{\Delta }(x,x^{\prime };q)=\sgn q_{1}\delta ^{\prime
}(x_{2}^{{}}-x_{2}^{\prime })\sum_{n=1}^{N}\vartheta _{n}(q_{1})\widetilde{\Phi }_{n}(x,q_{1})\widetilde{\Psi }_{n}(x^{\prime },q_{1}).  \label{hLd}
\end{equation}
Now it is easy to check directly that
\begin{equation*}
PL_{\Delta }=L_{\Delta }P=L_{\Delta },
\end{equation*}
as it also follows from from~(\ref{delta}) and~(\ref{P2}).

In Sec.~\ref{Resolvent} we proved that in order to obtain $M_{\Delta }$ it
is enough to find, first, a general (belonging to our space of operators)
solution $X$ of the system (\ref{X}) and then to find a special $X$
satisfying (\ref{LDMD}).

The expression for $\widehat{L}_{\Delta }$ in~(\ref{hLd}) suggests that the
general solution $X$ of~(\ref{X}) is given by
\begin{equation}
\widehat{X}(x,x^{\prime };q)=\sum_{n=1}^{N}\vartheta
_{n}(q_{1})g_{m}(x_{2},x_{2}^{\prime };q)\widetilde{\Phi }_{n}(x,q_{1})\widetilde{\Psi }_{n}(x^{\prime },q_{1}),  \label{hX}
\end{equation}
with $g_{m}(x_{2},x_{2}^{\prime };q)$ such that $X(x,x^{\prime };q)\in \mathcal{S}^{\prime }$, but otherwise arbitrary function of the written arguments. This
can be verified directly by using the representation of $\widehat{P}$ and
the orthogonality~(\ref{scalar+}).

Thus we take $\widehat{M}_{\Delta }$ in the form~(\ref{hX}) and, according
to the discussion at the end of Sec.~\ref{Relation}, we fix the unknown
function by requiring that equations~(\ref{LDMD}) for $X=M_{\Delta }$ are
satisfied. We get
\begin{equation}
\fl\widehat{M}_{\Delta }(x,x^{\prime };q)=-\sgn q_{1}\sum_{n=1}^{N}\vartheta
_{n}(q_{1})\{\theta (x_{2}-x_{2}^{\prime })+g_{m}(q)\}\widetilde{\Phi }_{n}(x,q_{1})\widetilde{\Psi }_{n}(x^{\prime },q_{1}),  \label{hMD}
\end{equation}
with $g_{m}(q)$ to be chosen in such a way that $M_{\Delta }(x,x^{\prime
};q) $ is bounded at space infinity. It results that
\begin{eqnarray}
\fl M_{\Delta }(x,x^{\prime };q) =-\sgn q_{1}\sgn(x_{2}^{{}}-x_{2}^{\prime
})e^{-q(x-x^{\prime })}  \nonumber \\
\times \sum_{n=1}^{N}\vartheta _{n}(q_{1})\theta ((q_{2}-2\lambda _{n\Re
}q_{1})(x_{2}^{{}}-x_{2}^{\prime }))\widetilde{\Phi }_{n}(x,q_{1})\widetilde{\Psi }_{n}(x^{\prime },q_{1})  \label{MD1}
\end{eqnarray}
obtained for
\begin{equation}
g_{m}(q)=-\theta (-q_{2}+2\lambda _{m\Re }q_{1})  \label{choice}
\end{equation}
is a well defined bounded operator.

Since the $\widetilde{\Phi }$'s and $\widetilde{\Psi }$'s are bounded we
need only to consider the region of variables where the exponent $e^{-q(x-x^{\prime })}$ is growing. Since, thanks to the choice~(\ref{choice}) in each term in the r.h.s.\ of~(\ref{hMD})
\begin{equation*}
-q_{2}(x_{2}-x_{2}^{\prime })\leq -2\lambda _{n\Re
}q_{1}(x_{2}-x_{2}^{\prime })
\end{equation*}
the behaviour of $e^{-q(x-x^{\prime })}$ for $q_{1}>0$ cannot be worse than $e^{-q_{1}((x_{1}-x_{1}^{\prime })+2\lambda _{n\Re }(x_{2}-x_{2}^{\prime }))}$
and, then, since $|q_{1}|\leq |\lambda _{n\Im }|$, no worse than $\theta
(\pm \lambda _{n\Im })e^{\mp \ell _{\Im }(\lambda _{n})(x-x^{\prime })}$.
But in the interval $0\leq q_{1}\leq |\lambda _{n\Im }|$ where $\Phi _{n}$
and $\Psi _{n}$ are different from zero
\begin{eqnarray*}
\fl\theta (\pm \lambda _{n\Im }) e^{\mp \ell _{\Im }(\lambda_{n})(x-x^{\prime
})}\widetilde{\Phi }_{n}(x,q_{1})\widetilde{\Psi }_{n}(x^{\prime },q_{1})\\
{ }=\theta (\pm \lambda _{n\Im })e^{-i\ell _{\Re }(\lambda _{n})(x-x^{\prime
})}\widetilde{\chi }(x,\lambda _{n_{\Re }}+i|\lambda _{n\Im }|)\widetilde{\xi }(x^{\prime },\lambda _{n_{\Re }}+i|\lambda _{n\Im }|)
\end{eqnarray*}
is bounded. Analogously for $q_{1}<0$. This proves that $M_{\Delta }$ is
bounded.

We are left with the explicit construction of $\zeta M\zeta ^{\dag }$. Form (\ref{zetanuom}) and the conjugation properties of the Jost solutions we have
\begin{equation*}
\zeta M\zeta ^{\dag }=\widetilde{\nu }\tau \omega M\nu \tau ^{\dag }\widetilde{\omega }
\end{equation*}
and, then, from (\ref{LM}), (\ref{scalar}) and (\ref{Tt})
\begin{equation*}
\zeta M\zeta ^{\dag }=\widetilde{\nu }TM_{0}\widetilde{\omega }.
\end{equation*}
In the $p$-space this reads
\begin{equation*}
\fl(\zeta M\zeta ^{\dag })(p;\mathbf{q})=\int dp^{\prime }\widetilde{\nu }(p-p^{\prime
};\mathbf{q}_{1}+p_{1}^{\prime })\frac{t(\mathbf{q}_{1}+p_{1}^{\prime })}{\mathbf{q}_{2}+p_{2}^{\prime }-(\mathbf{q}_{1}+p_{1}^{\prime })^{2}}\widetilde{\omega }(p^{\prime };\mathbf{q}_{1})
\end{equation*}
and shifting $p^{\prime }$, that is naming $p^{\prime }+\mathbf{q}_{\Re }=\alpha $,
\begin{equation*}
\fl(\zeta M\zeta ^{\dag })(p;\mathbf{q})=\int d\alpha \,\widetilde{\nu }(p-\alpha +\mathbf{q}_{\Re };\alpha _{1}+iq_{1})\frac{t(\alpha _{1}+iq_{1})}{\alpha
_{2}+iq_{2}-(\alpha _{1}+iq_{1})^{2}}\widetilde{\omega }(\alpha -\mathbf{q}_{\Re };\mathbf{q}_{1}).
\end{equation*}
Recalling (\ref{nuchi}) and inverting (\ref{xp}) we derive for $\zeta M\zeta
^{\dag }$ in the $x$-space
\begin{equation*}
\fl(\zeta M\zeta ^{\dag })(x,x^{\prime };q)=\frac{1}{(2\pi )^{2}}\int d\alpha
\frac{e^{-i\alpha (x-x^{\prime })}t(\alpha _{1}+iq_{1})}{\alpha
_{2}+iq_{2}-(\alpha _{1}+iq_{1})^{2}}\widetilde{\chi }(x;\alpha _{1}+iq_{1})\widetilde{\xi }(x^{\prime };\alpha _{1}+iq_{1}),
\end{equation*}
and, finally, integrating over $\alpha _{2}$, and summing up $M_{\Delta }$ as indicated in (\ref{MD1}) we get for the hat version of the resolvent $\widehat{M}$
\begin{eqnarray}
\fl\widehat{\widetilde{M}}(x,x^{\prime };q) =\sgn(x_{2}-x_{2}^{\prime })  \nonumber \\
\fl\quad\times\frac{1}{2\pi i}\!\!\int\!\! d\alpha _{1}\theta \left(
(q_{2}-2\alpha _{1}q_{1})(x_{2}-x_{2}^{\prime })\right)t(\alpha _{1}+iq_{1})\widetilde{\Phi }(x;\alpha _{1}+iq_{1})\widetilde{\Psi }(x^{\prime };\alpha _{1}+iq_{1})\nonumber\\
\fl\quad{ }-\sgn q_{1}\sgn(x_{2}^{{}}-x_{2}^{\prime
})\sum_{n=1}^{N}\vartheta _{n}(q_{1})\theta ((q_{2}-2\lambda _{n\Re
}q_{1})(x_{2}^{{}}-x_{2}^{\prime }))\widetilde{\Phi }_{n}(x,q_{1})\widetilde{\Psi }_{n}(x^{\prime },q_{1}).
\end{eqnarray}

It is worthwhile to note that in the r.h.s.\ of this formula for $\widetilde{M}$
the poles of $t(\mathbf{q}_{1})$ at $\mathbf{q}_{1}=\lambda _{n}$ and at $\mathbf{q}_{1}=\overline{\lambda }_{n}$ in the first term generate discontinuities at $q_{1}=\pm \lambda _{n\Im }$ which
cancel exactly the discontinuities of the second term, due to
the fact that the auxiliary Jost solutions $\Phi _{n}(q_{1})$, $\Psi
_{n}(q_{1})$ are identically zero outside the strip $|q_{1}|\leq |\lambda
_{n\Im }|$.

By using the reduction procedures indicated in (\ref{Gk}) and (\ref{Gpm}) from this explicit expression for the resolvent one can derive the generalization of the standard Green's function for Jost and advanced/rearted solutions to the case of $N$ solitons superimposed to a generic background. Thanks to the fact that they are derived from the resolvent they result both to be bilinear in terms of the Jost and auxiliary Jost solutions. The use of these Green's functions and other ones suggested by studying the properties of the resolvent is crucial in extending the IST to potentials obtained by perturbing the potential considered in this paper by adding to it an arbitrary smooth rapidly decaying function of both spatial variables. This is performed in a forthcoming paper following the method developed in \cite{1-line}.

\ack This work is supported in part by the Russian Foundation for Basic Research (grant \# 05-01-00498), Scientific Schools 2052.2003.1., and by the Program of RAS ``Mathematical Methods of the Nonlinear Dynamics.'', by PRIN 2004 `Sintesi' and by INFN. The authors acknowledge fruitful discussions
with Dr. Barbara Prinari. A.~K.~P. thanks his colleagues
in the Department of Physics of Lecce University for kind hospitality.

\end{document}